\documentclass[12pt]{spieman}
\usepackage[margin=1in]{geometry}

\usepackage{amsmath,amsfonts,amssymb}
\usepackage{graphicx}
\usepackage{setspace}
\usepackage{tocloft}
\usepackage{lineno}
\usepackage{booktabs}
\usepackage{array}
\usepackage{tabularx}
\usepackage{multirow}
\usepackage{makecell}
\usepackage{threeparttable}
\usepackage{microtype}
\usepackage{xcolor}
\usepackage{enumitem}
\usepackage{url}
\usepackage{seqsplit}
\usepackage{mathtools}
\usepackage{float}
\usepackage{longtable}
\usepackage{orcidlink}
\usepackage{caption}

\newcommand{\ind}{\mathbb{1}}

\title{Bayesian Uncertainty Quantification of Breast-Cancer Pathology AI Agreement under Reader- and Annotation-Condition-Dependent Reference Labels}

\author{Hammed Olayinka \orcidlink{0000-0002-9796-5276}}
\affil{Department of Mathematical Sciences, Worcester Polytechnic Institute, 100 Institute Road, Worcester, Massachusetts 01609, United States}

\begin{document}
\maketitle

\begin{abstract}
\textbf{Purpose:} Human reference labels used to evaluate medical artificial intelligence (AI) can vary across readers and annotation workflows. This study quantifies agreement of one fixed pathology AI system with a future qualified reader without treating an inferred consensus as error-free biological truth, and separates ordinary sampling uncertainty from uncertainty induced by the reference process itself.

\textbf{Approach:} This study analyzed linked Evaluation and Unbiased multirater data from NuCLS using a fixed external HoVer-Net decision vector. The primary selected frame contained 1,144 anchors, 52 fields of view, five slides/patients, and five primary readers. The target was fixed-AI agreement with a future reader under condition $c$, $\theta_c^S=P(A_i=Y_{i,r^*,c}\mid S_i=1)$, with descriptive contrast $\Delta^S=\theta_E^S-\theta_U^S$. The analysis used raw agreement and iMRMC benchmarks, a reader-only Bayesian model, a working shared-detectability model, residual-dependence and nonignorable-missingness sensitivities, slide and future-reader sensitivities, frozen adversarial simulations, and nested predictive assessment with cluster-aware uncertainty.

\textbf{Results:} Among 5,188 reader-anchor pairs observed in both conditions, 19.62\% changed AI-agreement state, yet raw agreement was 0.5466 in Unbiased and 0.5473 in Evaluation. The reader-only model gave $\Delta^S=-0.0008$ (95\% credible interval, $-0.0033$ to $0.0008$), and the shared-detectability model gave $\Delta^S\approx0$ ($-0.0140$ to $0.0127$). Across structural perturbations, point estimates of absolute agreement ranged from 0.420 to 0.646 in Unbiased and from 0.409 to 0.640 in Evaluation. In frozen adversarial simulations, the prespecified comparative-stability criterion passed (median $R=0.295$; 14/16 model-scenario cells had $R<1$), with condition-specific AI-human dependence as the common robustness boundary. Held-out-AI prediction supported shared predictive information within the selected frame, but predictive behavior varied materially by slide composition.

\textbf{Conclusions:} In this selected NuCLS frame, the average Evaluation-minus-Unbiased agreement contrast was comparatively stable across many analyses, while absolute fixed-AI/future-reader agreement was structurally sensitive. This does not establish equivalence, causal assistance effects, biological accuracy, or new-patient transportability. Medical-imaging AI evaluations using variable human references should report reader population, selected frame, observation mechanism, and structural sensitivity alongside conventional agreement summaries.
\end{abstract}

\keywords{Computational pathology, Multi-reader multi-case, Reference standard, Bayesian hierarchical modeling, Human--AI agreement}

{\noindent\footnotesize Correspondence: \linkable{haolayinka@wpi.edu}\par}

\begin{spacing}{1}

\section{Introduction}

Medical-imaging AI systems are commonly evaluated against a reference standard that is operationally treated as the target label. Yet many clinical reference standards are generated by expert interpretation rather than by an error-free measurement process. Current reporting guidance explicitly recommends the term ``reference standard'' rather than ``ground truth'' when labels can be uncertain, and it emphasizes documenting annotator qualifications, interobserver variability, missingness, and robustness analyses.\cite{Tejani2024,Hadjiiski2023,Hanna2024} Work on AI validation and expert-panel reference standards likewise emphasizes that the reference procedure itself is a design choice whose composition and adjudication can affect measured performance.\cite{Duggan2021,Kellerhuis2025} In computational pathology, annotation guidance documents substantial inter-pathologist variability, missed objects, and the dependence of annotations on definitions and workflow.\cite{Wahab2022} Breast-pathology reader studies have also demonstrated substantial diagnostic disagreement even when interpretations are compared with a consensus-derived reference.\cite{Elmore2015} A recent breast-pathology AI reader study reported that measured AI accuracy was highly dependent on reference-panel composition.\cite{Shamshoian2026}

Methods for imperfect references are not new. Reviews of diagnostic evaluation without gold standards predate modern medical AI,\cite{HuiZhou1998} Dawid--Skene models estimate annotator error rates without an observed truth label,\cite{DawidSkene1979} STAPLE jointly estimates a probabilistic segmentation reference and rater performance,\cite{Warfield2004} and latent-class approaches estimate test characteristics when a perfect gold standard is unavailable.\cite{CollinsHuynh2014} These models remain assumption-dependent: latent-class models can overstate accuracy under some misspecifications,\cite{Spencer2012} conditional-dependence structure can materially alter inferred sensitivity and specificity,\cite{Keddie2023} and dependent errors between a test and an imperfect reference can bias apparent performance in either direction.\cite{Walter2012} Contemporary image-analysis work also models multiple annotators explicitly rather than collapsing them immediately to one label.\cite{Zhang2023,Jalal2026} Partial-identification approaches provide another route when assumptions are deliberately weakened,\cite{Obradovic2024} but they answer a different question from the structural-sensitivity analysis developed here.

A parallel multireader multicase (MRMC) literature addresses uncertainty induced by sampling readers and cases. Binary-agreement MRMC methodology is established,\cite{Chen2014} as are designs and analyses intended to generalize across reader and case populations when those samples are representative.\cite{Obuchowski2022} The FDA iMRMC tool provides current regulatory-science software for MRMC analysis, including binary-data functionality,\cite{FDA2025} and recent work continues to develop analysis for binary medical tests in MRMC designs.\cite{LeeMRMC2026} Population-based generalized mixed models and Bayesian random-effects approaches likewise predate this study.\cite{Nelson2010,Hsiao2011} Recent Bayesian MRMC work explicitly addresses small reader samples, reader heterogeneity, case effects, and reader-case interactions.\cite{Jacobs2026} Consequently, neither random-reader inference nor Bayesian hierarchical multirater modeling is claimed here as a new statistical primitive.

The additional complication in the present setting is that human labels may depend on the annotation workflow. Human--AI interaction studies show that readers can alter decisions in response to algorithmic suggestions, including incorrect suggestions, and that AI assistance can change both diagnostic decisions and interreader agreement.\cite{Dratsch2023,LeeHumanAI2023,Fari2025} NuCLS is particularly useful for examining this issue because its multirater annotations were collected under conditions that differed in exposure to algorithmic suggestions.\cite{Amgad2022} The Unbiased control was collected without algorithmic suggestions, whereas the Evaluation condition showed refined algorithmic suggestions. NuCLS also used a Dawid--Skene-style expectation-maximization procedure to infer pathologist-derived P-truth labels. That inferred reference is a model-based multirater construct, not independently verified biological truth. Because the Unbiased condition preceded the suggestion-assisted conditions, Evaluation-versus-Unbiased comparisons are also entangled with order and experience and cannot be interpreted as randomized causal effects of assistance.\cite{Amgad2022}

This study therefore uses a different target. Let $A_i$ denote the fixed binary decision of one external AI system for selected anchor $i$, let $Y_{i,r^*,c}$ denote the response of a future qualified reader under annotation condition $c\in\{U,E\}$, and let $S_i=1$ denote inclusion in the linked selected frame. The condition-specific estimand is
\begin{equation}
\theta_c^S=P\!\left(A_i=Y_{i,r^*,c}\mid S_i=1\right),
\label{eq:estimand}
\end{equation}
with descriptive contrast
\begin{equation}
\Delta^S=\theta_E^S-\theta_U^S.
\label{eq:contrast}
\end{equation}
This target treats reader-population variability as part of agreement uncertainty rather than replacing the readers by a single assumed truth label.

\begin{figure}[H]
\centering
\includegraphics[width=\textwidth]{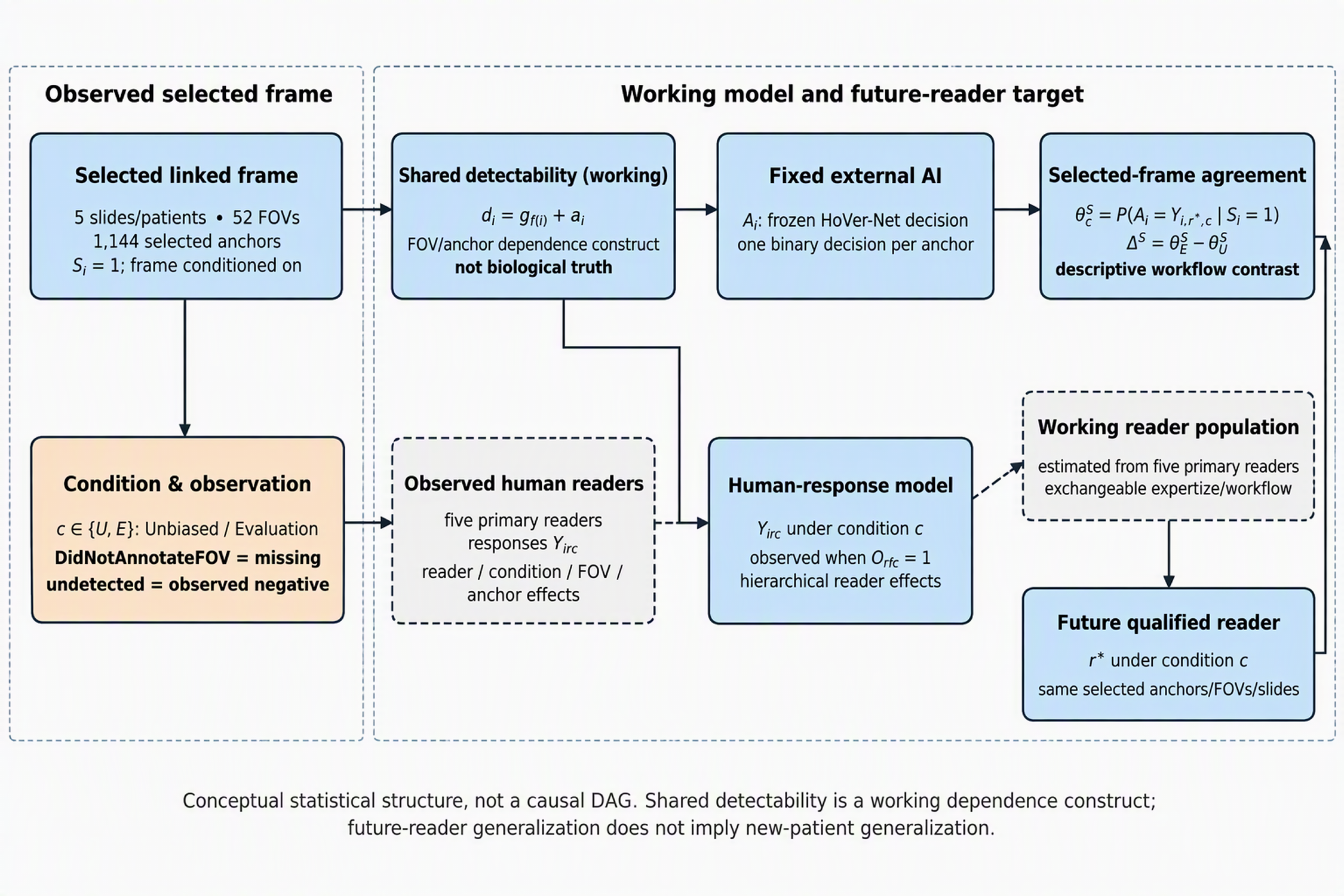}
\caption{\label{fig:conceptual}Conceptual framework for selected-frame fixed-AI/future-reader agreement. The selected NuCLS frame feeds a working shared-detectability structure used by the fixed external AI channel and the human reader process. Reader-population variation is integrated to define agreement with a future qualified reader. The diagram is a statistical framework, not a causal DAG: shared detectability is not biological truth, \texttt{DidNotAnnotateFOV} is missing rather than negative, and the Evaluation--Unbiased contrast is descriptive.}
\end{figure}

The contribution is therefore an estimand-and-evidence framework rather than a claim of generic multirater-model novelty. This study combines direct empirical agreement, conventional MRMC benchmarking, Bayesian future-reader integration, explicit structural sensitivities, frozen adversarial misspecification experiments, and predictive calibration to distinguish conclusions that remain comparatively stable from quantities that are structurally fragile. Figure~\ref{fig:conceptual} summarizes the inferential pathway. The left side identifies the linked selected NuCLS frame and the fixed external AI decision for each anchor. The center depicts the working FOV- and anchor-level structure used to represent common variation in detectability, while the human-response branch adds reader, reader-by-condition, reader-by-FOV, and reader-by-anchor variation. The right side shows that these components are integrated over a future qualified reader to obtain condition-specific agreement. The figure is conceptual: it does not assert biological truth, establish a causal pathway, or imply that unobserved FOVs are negative observations.

\section{Materials and Methods}

\subsection{NuCLS multirater data and fixed external AI}

NuCLS contains single- and multirater nucleus annotations from breast-cancer histopathology images under Evaluation, Bootstrap-control, and Unbiased-control conditions.\cite{Amgad2022} The present study uses the linked Evaluation and Unbiased multirater data. Five primary readers were retained: SP.3, JP.1, JP.2, JP.5, and JP.6. SP.2 was not included in the primary reader population because paired support was limited to one slide, so its inclusion would require extrapolating a reader-condition contribution from single-slide support.

The final linked frame contained 1,144 selected anchors across 52 FOVs and five slides/patients. The frozen external AI vector contained one HoVer-Net\cite{Graham2019} binary decision per anchor, with 593 AI-positive and 551 AI-negative anchors. This external HoVer-Net was distinct from the suggestion-generating system used in the original NuCLS annotation workflow. The study therefore evaluates agreement between one fixed external AI and readers observed under different NuCLS workflows; it does not estimate the effect of HoVer-Net assistance.

For five readers, two conditions, and 1,144 selected anchors, the potential reader-anchor-condition frame contained 11,440 records. Of these, 10,833 were observed. NuCLS distinguishes \texttt{undetected} from \texttt{DidNotAnnotateFOV}; the former was encoded as an observed negative response, whereas the latter was treated as missing and never recoded as a negative outcome. In the

\begin{table}[H]
\centering
\setlength{\tabcolsep}{20pt}
\begin{threeparttable}
\caption{Summary statistics}
\label{tab:frame} 
\footnotesize
\begin{tabular}{lrr}
\toprule
Quantity & Count & Percent \\
\midrule
Selected anchors & 1,144 & -- \\
AI-positive anchors & 593/1,144 & 51.84\% \\
AI-negative anchors & 551/1,144 & 48.16\% \\
Fields of view & 52 & -- \\
Slides/patients & 5 & -- \\
Primary readers & 5 & -- \\
Potential reader-anchor-condition records & 11,440 & -- \\
Observed records & 10,833/11,440 & 94.69\% \\
Unobserved because reader did not annotate FOV & 607/11,440 & 5.31\% \\
\midrule
Evaluation missing & 516/5,720 & 9.02\% \\
\quad AI-negative anchors & 271/2,755 & 9.84\% \\
\quad AI-positive anchors & 245/2,965 & 8.26\% \\
Unbiased missing & 91/5,720 & 1.59\% \\
\quad AI-negative anchors & 80/2,755 & 2.90\% \\
\quad AI-positive anchors & 11/2,965 & 0.37\% \\
\midrule
Potential reader-anchor pairs & 5,720 & -- \\
Observed in both conditions & 5,188/5,720 & 90.70\% \\
Agree with AI in both conditions & 2,326/5,188 & 44.83\% \\
Agree with AI in neither condition & 1,844/5,188 & 35.54\% \\
Agree only in Evaluation & 515/5,188 & 9.93\% \\
Agree only in Unbiased & 503/5,188 & 9.70\% \\
Switch AI-agreement state & 1,018/5,188 & 19.62\% \\
\bottomrule
\end{tabular}
\begin{tablenotes}[flushleft]
\scriptsize
\item \textit{Note.} The table displays empirical support, observation structure, and paired agreement states in the primary five-reader frame. Percentages use the denominator shown in the count column. \texttt{DidNotAnnotateFOV} denotes an unobserved response, whereas \texttt{undetected} denotes an observed negative response.
\end{tablenotes}
\end{threeparttable}
\end{table}

\noindent verified frame, the observation mask was constant within each reader-by-FOV-by-condition cell: no FOV showed sporadic within-FOV missingness. Thus, \texttt{DidNotAnnotateFOV} was treated as a reader-by-FOV-by-condition observation process rather than as anchor-level non-detection. A fully crossed subset contained 679 anchors, 32 FOVs, four slides, and all five primary readers observed under both conditions for each anchor. Table~\ref{tab:frame} summarizes the selected-frame support, observation structure, missingness, and paired agreement states used throughout the analysis.

\subsection{Agreement target and scope of inference}

The analysis separates three quantities that are often combined in AI evaluation. The observed data describe how the five participating readers responded to the selected anchors. The statistical models describe the distribution of responses expected from a new qualified reader resembling those participants. The fixed external AI contributes one frozen binary decision for each anchor. The primary outcome is whether that fixed AI decision matches the response generated for the future reader; it is an agreement outcome, not a claim that either party is biologically correct.

Equation~\eqref{eq:estimand} generalizes over a future reader, not over new patients or a new AI system. The reader population is interpreted as qualified readers plausibly exchangeable with the five primary readers in expertise and workflow. This exchangeability is a working model assumption, not an empirically verified probability sample of a broader pathologist population. The 1,144 anchors, 52 FOVs, five slides/patients, and fixed HoVer-Net vector are conditioned on in the primary estimand. Slide-hierarchy and leave-one-slide-out analyses are therefore composition sensitivities, not new-patient validation. The condition contrast is descriptive rather than causal.

No equivalence margin was specified externally. Accordingly, intervals containing zero are not interpreted as evidence that the two conditions are equivalent. This study uses ``near zero within observed precision'' when describing small contrasts.

\subsection{Analysis framework}

The analysis proceeded from direct summaries to progressively more structured models. M0 comprised raw agreement and iMRMC benchmarks and therefore did not impose a latent reference-process model. M1 modeled variation among readers, FOVs, and anchors without using the AI decisions in the human-response likelihood. M2 added a shared-detectability term as a working representation of information common to readers and the fixed AI. M3 then relaxed the conditional-independence assumption by allowing residual AI--human association. This sequence was designed to show which conclusions were supported by direct observation and which depended on additional reference-process assumptions.

\subsubsection{Reader-response model}

For anchor $i$ in FOV $f(i)$, define the working FOV/anchor structure
\begin{equation}
d_i=g_{f(i)}+a_i,
\label{eq:di}
\end{equation}
where $g_f$ and $a_i$ are centered Gaussian FOV and anchor effects. For reader $r$, condition $c\in\{U,E\}$, and an observed record,
\begin{align}
D_{irc} &\sim \mathrm{Bernoulli}(p_{irc}),\\
\mathrm{logit}(p_{irc}) &= \alpha+\beta_E\ind(c=E)+d_i+u_r+b_r\ind(c=E)+h_{r,f(i)}+q_{ri}.
\label{eq:reader}
\end{align}
Here $u_r$ is a baseline reader effect, $b_r$ is a reader-specific condition deviation, $h_{rf}$ is a reader-by-FOV effect, and $q_{ri}$ is a reader-by-anchor effect. Baseline priors were $\alpha,\alpha_A\sim N(0,2.5^2)$, $\beta_E\sim N(0,1)$, $\lambda_A\sim\mathrm{HalfNormal}(1.5)$, $\sigma_r,\sigma_{rE},\sigma_f,\sigma_{rf}\sim\mathrm{HalfNormal}(1)$, and $\sigma_a,\sigma_{ra}\sim\mathrm{HalfNormal}(1.5)$. Noncentered parameterizations were used for Gaussian random effects.

M1 used Equation~\eqref{eq:reader} only. The fixed AI vector was not included in the human-response likelihood; it entered only after fitting when reader probabilities were transformed into AI-reader agreement probabilities.

\subsubsection{Shared-detectability working model}

M2 added the fixed external AI channel
\begin{equation}
A_i\sim\mathrm{Bernoulli}\left[\mathrm{logit}^{-1}(\alpha_A+\lambda_A d_i)\right],\qquad \lambda_A>0.
\label{eq:ai}
\end{equation}
The reader loading on $d_i$ is fixed to one, defining the latent scale relative to the reader logit, and the positive constraint on $\lambda_A$ defines orientation. These are parameterization constraints, not proof that the latent decomposition is empirically unique. The shared factor is a working dependence representation and is not interpreted as biological truth. The likelihood factorization, latent-scale and orientation constraints, and derivation of the selected-frame future-reader agreement estimand are provided in the Supplementary Material.

\subsection{Future-reader integration}

To translate the fitted models from the five observed readers to the stated reader-population target, each posterior draw generated a new reader with new reader-level and reader-by-context effects. In plain terms, this step averages over how qualified readers may differ rather than treating any participating reader or consensus vote as the reference. In the full future-reader construction, the new reader receives new realizations of the reader-level, reader-by-condition, reader-by-FOV, and reader-by-anchor effects. Under the baseline independent reader-effect parameterization, the marginal future-reader variance is
\begin{align}
\sigma_U^2 &= \sigma_r^2+\sigma_{rf}^2+\sigma_{ra}^2,\\
\sigma_E^2 &= \sigma_r^2+\sigma_{rE}^2+\sigma_{rf}^2+\sigma_{ra}^2.
\end{align}
For posterior draw $m$, future-reader detection probabilities were obtained as logistic-normal expectations,
\begin{align}
p_{iU}^{*(m)} &= E_Z\left[\mathrm{logit}^{-1}\{\alpha^{(m)}+d_i^{(m)}+\sigma_U^{(m)}Z\}\right],\\
p_{iE}^{*(m)} &= E_Z\left[\mathrm{logit}^{-1}\{\alpha^{(m)}+\beta_E^{(m)}+d_i^{(m)}+\sigma_E^{(m)}Z\}\right],
\end{align}
with $Z\sim N(0,1)$ and numerical integration by Gauss--Hermite quadrature. Conditional on the fixed AI decision,
\begin{equation}
q_{ic}^{(m)}=A_i p_{ic}^{*(m)}+(1-A_i)\{1-p_{ic}^{*(m)}\},
\end{equation}
and $\theta_c^S$ is the selected-anchor average of $q_{ic}$.

A new reader-by-anchor draw on a fixed observed anchor represents reader-specific perception or interpretation for that same nucleus; it does not create a new anchor or a new biological state. Because this population construction is not directly verifiable, this study also calculated an alternative future-reader estimand that retained the observed-reader likelihood but omitted new reader-by-anchor variance from future-reader integration.

Baseline reader tendency and reader-specific condition response may be correlated. With only five primary readers, that population correlation is weakly supported by the data. Rather than estimate it precisely, this study refit M1 and M2 over a fixed sensitivity grid $\rho_{ub}\in\{-0.90,-0.75,\ldots,0.90\}$ using a correlated bivariate reader-effect construction with preserved marginal scales.

The logistic-normal reduction, Gauss--Hermite approximation, correlated-reader variance calculation, and alternative future-reader constructions are detailed in the Supplementary Material; Supplementary Table S1 summarizes the resulting construction sensitivity.

\subsection{Structural and observation-process sensitivities}

\subsubsection{Residual AI--human dependence sensitivity (M3)}

M3 starts from the reader-only future-reader probability $p^0_{ic}$ and imposes a fixed residual association with the observed AI decision,
\begin{equation}
\mathrm{logit}(p^\gamma_{ic})=\mathrm{logit}(p^0_{ic})+\kappa_c(\gamma)+\gamma A_i,
\label{eq:m4}
\end{equation}
where $\kappa_c(\gamma)$ preserves mean future-reader positivity within each posterior draw and condition. The analysis evaluated $\gamma\in[-2,2]$ in increments of 0.25 (odds-ratio range approximately 0.135 to 7.39). M3 is a sensitivity grid, not an oracle-selected estimator and not a formal partial-identification bound.

\subsubsection{Nonignorable-missingness pattern-mixture sensitivity}

At each FOV/condition, future-reader prediction was represented as a mixture of an observed-like regime and a \texttt{DidNotAnnotateFOV}-like regime. The mixing weight was the empirical fraction of the five primary readers missing that FOV/condition, and the missing-like log odds were shifted by $\gamma_{\mathrm{mis}}\in[-2,2]$ in increments of 0.25. The observed-data posterior was held fixed. This delta-adjusted posterior-predictive calculation is a sensitivity analysis, not an identified missingness mechanism. At $\gamma_{\mathrm{mis}}=0$, it reproduced the fitted M1/M2 summaries to numerical precision.

\subsubsection{Slide hierarchy and slide-composition sensitivity}

A slide/patient effect was added above the FOV and anchor effects in M2, with $\sigma_{\mathrm{slide}}\sim\mathrm{HalfNormal}(0.5)$ and noncentered parameterization. Because only five slides/patients were observed, the slide-scale posterior is interpreted cautiously. Five leave-one-slide-out (LOSO) refits were used to assess composition sensitivity. These refits are not external validation and do not establish new-patient transportability. Full residual-dependence, pattern-mixture missingness, and leave-one-slide-out specifications and results are reported in Supplementary Tables S4--S6 and Supplementary Figure S1.

\subsection{Identification and prior sensitivity}

The M2 scale is mathematically anchored by fixing the human loading to one, centering random effects, and restricting $\lambda_A>0$. Empirical information is different. Repeated human responses across readers, conditions, FOVs, and anchors inform the reader and shared FOV/anchor structure. The AI intercept $\alpha_A$ is informed primarily by the prevalence of positive decisions in the fixed 1,144-anchor AI vector conditional on $d_i$, while $\lambda_A$ is informed by whether cross-anchor AI variation aligns with the shared structure inferred jointly from repeated human responses. Because there is only one binary AI observation per anchor, the data do not independently replicate anchor-level AI outcomes.

This study therefore examined prior-to-posterior concentration, weaker and more-regularizing prior families, and posterior dependence among $\lambda_A$, $\sigma_f$, $\sigma_a$, $\sigma_{rf}$, and $\sigma_{ra}$. Stable derived agreement estimands across those prior families are interpreted as prior robustness within the examined range, not as proof that every latent component is uniquely identified. Supplementary Tables S2 and S3 report the derived-estimand prior sensitivity and posterior parameter-correlation diagnostics.

\subsection{Adversarial simulation program}

The structural robustness program used one frozen set of 800 generated datasets: eight adversarial scenarios (A1--A8), 100 replicates each, generated once before model fitting and supplied identically to every applicable method. Truth files were separated from fitting code and were read only for evaluation. The scenarios challenged: (A1) independent AI-specific FOV/anchor variation; (A2) correlated but nonidentical AI/human factors; (A3) nonlinear/threshold AI response; (A4) condition-specific AI--human dependence; (A5) omitted slide/patient effects; (A6) informative FOV missingness; (A7) selection on AI/human positivity; and (A8) mixture/heavy-tailed anchor heterogeneity.

For each M1/M2 scenario cell, comparative stability was summarized by
\begin{equation}
R=\frac{\mathrm{RMSE}(\Delta^S)}{\{\mathrm{RMSE}(\theta_U^S)+\mathrm{RMSE}(\theta_E^S)\}/2}.
\label{eq:R}
\end{equation}
Before fitting, the qualitative conclusion ``contrast comparatively more stable than absolute level'' was prespecified to require median $R\le0.50$ across the 16 model-scenario cells and at least 12/16 cells with $R<1$. M3 was evaluated by truth containment across its full fixed sensitivity envelope rather than by oracle selection of $\gamma$. The exact scenario mechanisms, frozen truth-screen summaries, absolute-target coverage, Monte Carlo uncertainty calculations, and M3 envelope-containment results are provided in Supplementary Tables S7--S9.

\subsection{Predictive assessment}

Five outer folds held out anchors, not slides, retained all 52 FOVs and all five slides/patients, so the primary cross-validation is an observed-context new-anchor check rather than slide-disjoint or new-patient validation. Held-out-human prediction compared joint and reader-only Bayesian models. Held-out-AI prediction assessed whether human-response structure contained information about the fixed AI decision. Target-aligned prediction evaluated the observed-frame endpoint $\ind(A=D)$ directly.

For held-out AI, the latent Bayesian predictor was compared under the same outer folds with regularized logistic regression using reader-summary features and with histogram gradient boosting. Hyperparameter tuning and feature construction were performed inside each outer training fold. Performance summaries included Brier score, log loss, AUROC, AUPRC with event prevalence, calibration intercept and slope, and reliability curves. Calibration was emphasized because discrimination alone does not establish probabilistic reliability.\cite{VanCalster2019} Anchor-cluster and slide-cluster bootstrap intervals were calculated. Because only five slides were available, slide-bootstrap intervals and LOSO predictive analyses are interpreted as composition sensitivity rather than precise patient-population uncertainty. Nested tuning choices, condition-specific predictive metrics and reliability curves, and slide-disjoint predictive diagnostics are reported in Supplementary Tables S10--S12 and Supplementary Figures S2 and S3.

\subsection{Computation and software}

Bayesian models were fit in PyMC using four chains and noncentered parameterizations. Production settings varied by analysis, with high target acceptance for the more complex structural fits. Computational gates used divergent-transition counts, scalar $\widehat R$, and bulk/tail effective sample sizes. One prespecified M2 adversarial replication in A1 did not satisfy the computational gate after targeted retry; the corresponding summary therefore used 99 valid replications, as reported in the table note.

\subsection{Comparison of evaluation targets}

Table~\ref{tab:comparison} distinguishes the principal evaluation approaches according to their target quantity, treatment of the fixed AI output, representation of reader variation, and assumptions about the reference process. Raw observed agreement and binary iMRMC provide nonlatent empirical comparators, whereas M1 models the human-reader process without allowing the AI observations to inform model fitting. Latent-class and consensus approaches instead target an unobserved reference construct. M2 jointly represents the fixed AI and human responses through a shared-detectability structure, while the structural-sensitivity framework evaluates whether the resulting agreement levels and condition contrast remain stable under alternative defensible assumptions. These approaches therefore answer related but nonidentical questions, and differences among their estimates should be interpreted as consequences of their distinct targets and assumptions rather than as a ranking of methods.

\begin{table}[htpb]
\centering
\caption{Comparison of evaluation targets}
\label{tab:comparison}

\begin{threeparttable}
\footnotesize
\setlength{\tabcolsep}{3pt}
\renewcommand{\arraystretch}{0.98}

\noindent\parbox{\textwidth}{\raggedright\textbf{Panel A. Target and data role}}\\[1mm]
\begin{tabularx}{\textwidth}{@{}p{0.145\textwidth}>{\raggedright\arraybackslash}X>{\raggedright\arraybackslash}X>{\raggedright\arraybackslash}X@{}}
\toprule
Approach & Target quantity & AI status & Reader variation \\
\midrule
Raw observed agreement & Observed fixed-AI/reader concordance on the selected frame & Fixed external AI enters through observed agreement & Descriptive; no reader-population model \\
Binary iMRMC & Reader/case-averaged binary agreement and condition contrast & Fixed AI defines the agreement endpoint in this application & Explicit reader/case covariance and uncertainty \\
Reader-only model (M1) & Future-reader agreement after modeling human responses & AI excluded from the human-response likelihood; used after fitting & Hierarchical reader and interaction effects \\
Latent-class / consensus & Latent class/consensus and annotator/test error parameters & AI may be absent or treated as another test/annotator & Often annotator-specific error rates \\
Shared-detectability model (M2) & Fixed-AI/future-reader agreement under joint shared structure & One fixed binary AI observation per anchor enters jointly & Reader, reader$\times$FOV, reader$\times$anchor, FOV, and anchor effects \\
Structural-sensitivity framework & Stability/range of $\theta_U^S$, $\theta_E^S$, and $\Delta^S$ across assumptions & Same frozen AI vector across comparisons & Reader uncertainty propagated across alternative models/constructions \\
\bottomrule
\end{tabularx}

\vspace{1mm}
\textbf{Panel B. Reference-process assumptions and interpretation}
\begin{tabularx}{\textwidth}{@{}p{0.125\textwidth}>{\raggedright\arraybackslash}X>{\raggedright\arraybackslash}X>{\raggedright\arraybackslash}X>{\raggedright\arraybackslash}X@{}}
\toprule
Approach & Imperfect reference labels & Principal identifying assumption & Generalization / causal interpretation & Appropriate use / representative literature \\
\midrule
Raw observed agreement & No latent correction; no biological-truth claim & Recorded responses and observation denominator are empirical objects & Selected frame only; Evaluation--Unbiased is descriptive & Transparent empirical anchor; current reference-standard guidance\cite{Tejani2024,Hadjiiski2023,Duggan2021} \\
Binary iMRMC & No latent biological truth required & MRMC covariance/sampling structure; population inference requires appropriate reader/case sampling & Reader/case populations only under appropriate sampling; no causal interpretation by itself & Conventional nonlatent comparator\cite{Chen2014,Obuchowski2022,FDA2025,LeeMRMC2026} \\
Reader-only model (M1) & Models reader responses directly & Logistic-Gaussian hierarchy and future-reader exchangeability & Future-reader only; anchors/FOVs/slides fixed; condition contrast descriptive & Separates reader-process modeling from joint AI-human sharing\cite{Nelson2010,Hsiao2011,Jacobs2026} \\
Latent-class / consensus & Explicit latent reference construct & Dependence structure, constraints, priors, and identifiability & Model/design dependent; no causal interpretation implied & Appropriate when latent consensus/class is itself the target\cite{DawidSkene1979,Warfield2004,CollinsHuynh2014,Keddie2023} \\
Shared-detectability model (M2) & No error-free truth; shared detectability is a working dependence structure & Shared-factor specification, scale/orientation constraints, priors; decomposition not uniquely verified & Future reader on fixed selected frame; no causal condition effect & Working structural model; not a claim of new generic Bayesian MRMC \\
Structural-sensitivity framework & Explicitly varies reference-process assumptions & Sensitivity set must be scientifically defensible; no sharp bounds claimed & Selected-frame target; LOSO is composition sensitivity; no causal assistance effect & Distinguishes comparatively stable conclusions from structurally fragile levels\cite{Walter2012,Keddie2023,Obradovic2024} \\
\bottomrule
\end{tabularx}
\begin{tablenotes}[flushleft]
\scriptsize
\item \textit{Note.} M1, M2, and the structural-sensitivity framework are the implementations used in this study. The cited methods identify representative antecedent model families rather than exact equivalents. The comparison is descriptive and does not rank the approaches.
\end{tablenotes}
\end{threeparttable}
\end{table}

\section{Results}

\subsection{Observed annotation patterns and raw agreement}

Table~\ref{tab:frame} defines the empirical support. Of 11,440 potential reader-anchor-condition records, 10,833 (94.69\%) were observed. Missingness was substantially greater in Evaluation (9.02\%) than in Unbiased (1.59\%). Within Evaluation, missingness was 9.84\% for AI-negative anchors and 8.26\% for AI-positive anchors; within Unbiased, the corresponding rates were 2.90\% and 0.37\%. These differences are descriptive associations with the observed annotation workflow and are not interpreted causally.

Reader observation support was uneven but explicitly characterized. SP.3 contributed 1,069 observed Unbiased records and 846 Evaluation records; JP.1 contributed 1,144 and 1,120; JP.2 contributed 1,128 and 1,016; JP.5 contributed all 1,144 anchors in both conditions; and JP.6 contributed 1,144 and 1,078. At the FOV level, all five readers were observed in 48/52 Unbiased FOVs and 33/52 Evaluation FOVs. In Unbiased, the remaining four FOVs consisted of three FOVs missing SP.3 and one missing JP.2. In Evaluation, the remaining 19 FOVs consisted of 11 missing SP.3, four missing JP.2, two missing both SP.3 and JP.2, one missing SP.3 and JP.1, and one missing JP.6. No FOV showed within-FOV observation-mask variation for a given reader and condition, supporting the reader-by-FOV-by-condition observation mechanism described above.

Raw fixed-AI/reader agreement was nearly identical marginally: 2,848/5,204 (0.5473) in Evaluation and 3,077/5,629 (0.5466) in Unbiased, a difference of 0.00064 (0.064 percentage points). That marginal similarity concealed substantial paired instability. Among 5,188 reader-anchor pairs observed in both conditions, 2,326 (44.83\%) agreed with the AI in both and 1,844 (35.54\%) agreed in neither; 515 (9.93\%) agreed only in Evaluation and 503 (9.70\%) agreed only in Unbiased. Thus 1,018 pairs (19.62\%) switched AI-agreement state, while the imbalance between the two discordant directions was only 0.23 percentage points of paired observations. Directional reader-response switching was present in both AI strata: among AI-negative anchors, 243 paired responses changed from undetected in Unbiased to detected in Evaluation and 271 changed in the opposite direction, for a 20.82\% switch rate; among AI-positive anchors, the corresponding counts were 244 and 260, for an 18.54\% switch rate. These near-balanced directions illustrate how considerable individual label instability can cancel in the average agreement contrast.

\begin{figure}[htbp]
\centering
\includegraphics[width=\textwidth]{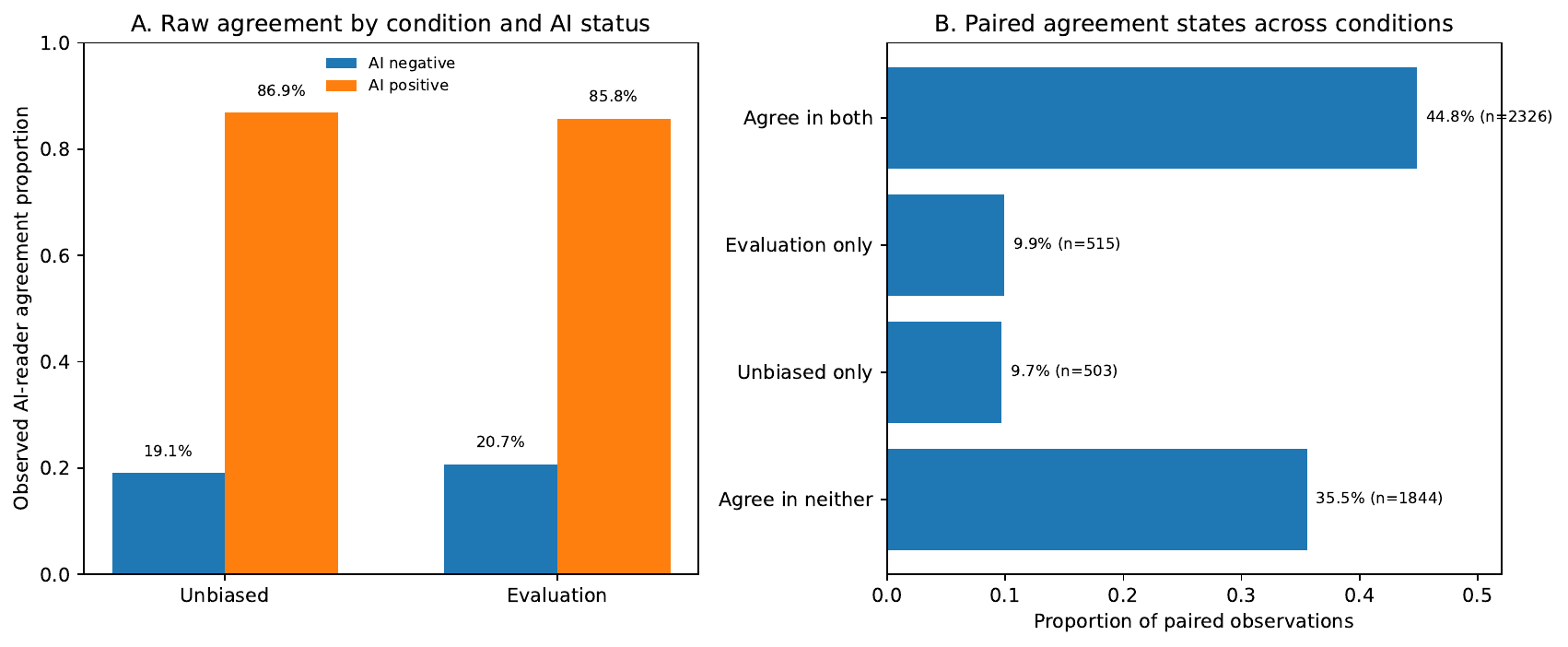}
\caption{\label{fig:raw}Observed fixed-AI/reader agreement across annotation conditions and paired condition-specific agreement states in the primary five-reader frame. Panel A stratifies raw observed agreement by annotation condition and fixed AI status. Panel B summarizes the 5,188 reader-anchor pairs observed in both conditions. The Evaluation-versus-Unbiased comparison is descriptive and does not estimate a causal effect of AI assistance.}
\end{figure}

Figure~\ref{fig:raw} separates two features that the overall means alone cannot show. Panel A stratifies observed agreement by fixed AI status: agreement for AI-positive anchors was 86.9\% in Unbiased and 85.8\% in Evaluation, whereas agreement for AI-negative anchors was 19.1\% and 20.7\%, respectively. Panel B displays the four paired agreement states and shows that the two discordant categories were almost balanced: 515 pairs agreed only in Evaluation and 503 only in Unbiased. Their near cancellation explains how 19.62\% switching could coexist with a marginal contrast of only 0.064 percentage points. The near-zero marginal condition contrast therefore should not be read as individual label stability.

\subsection{Agreement contrast across analysis frameworks}

The raw observed contrast was $\Delta^S=0.00064$. The corresponding five-reader iMRMC benchmark was 0.00064 with a 95\% confidence interval of $-0.02250$ to $0.02378$ under conditional-D and $-0.02394$ to $0.02521$ under joint-D. In the fully crossed 679-anchor subset, iMRMC estimated 0.00236 (95\% confidence interval, $-0.01405$ to $0.01876$).

The reader-only Bayesian model gave $\theta_U^S=0.5284$ (95\% credible interval, 0.5221 to 0.5348), $\theta_E^S=0.5276$ (0.5210 to 0.5343), and $\Delta^S=-0.0008$ ($-0.0033$ to $0.0008$), with posterior probability $P(\Delta^S>0)=0.160$. The shared-detectability working model gave $\theta_U^S=0.5787$ (0.5674 to 0.5908), $\theta_E^S=0.5787$ (0.5621 to 0.5956), and $\Delta^S\approx0$ ($-0.0140$ to $0.0127$). Adding a slide/patient effect changed the M2 contrast negligibly: $\Delta^S=-0.00004$ ($-0.01393$ to $0.01261$).

Across the five LOSO composition diagnostics, contrast point estimates ranged from $-0.00619$ to $0.00662$, with a combined interval envelope of $-0.02900$ to $0.03556$. Under M3 residual-dependence sensitivity, point estimates ranged from $-0.01109$ to $0.01010$, with a 95\% interval envelope of $-0.05238$ to $0.05127$. Pattern-mixture missingness sensitivity produced substantially smaller movement: the full-grid contrast span was 0.00341 for M1 and 0.00368 for M2. Correlating baseline reader tendency and condition response over $\rho_{ub}\in[-0.90,0.90]$ changed the full-construction contrast by only 0.084 percentage points in M1 and 0.076 percentage points in M2. The future-reader construction itself mattered more for M2: omitting new reader-by-anchor variation shifted the M2 contrast upward by approximately 0.24--0.39 percentage points across the correlation grid (0.278 percentage points at $\rho_{ub}=0$), whereas the corresponding M1 construction shift was smaller. All reported 95\% credible intervals for the contrast still included zero. The numerical sensitivity grids and leave-one-slide-out estimates underlying these ranges are provided in Supplementary Tables S1 and S4--S6 and summarized in Supplementary Figure S1.

Figure~\ref{fig:delta} places the conventional estimates and structural sensitivity analyses on one common Evaluation-minus-Unbiased scale. The raw and iMRMC rows provide nonlatent benchmarks; M1, M2, and the slide-hierarchy rows show model-based estimates and their inferential intervals; and the LOSO, residual-dependence, missingness, and future-reader rows show how the point estimate and interval envelope move across prespecified settings. Most point estimates remain close to zero, while the widest envelopes occur under slide-composition and residual-dependence perturbations. The repeated proximity of the point estimates to zero is evidence of comparative stability in this selected frame; overlap with zero is not an equivalence result.

\begin{figure}[htbp]
\centering
\includegraphics[width=0.8\textwidth]{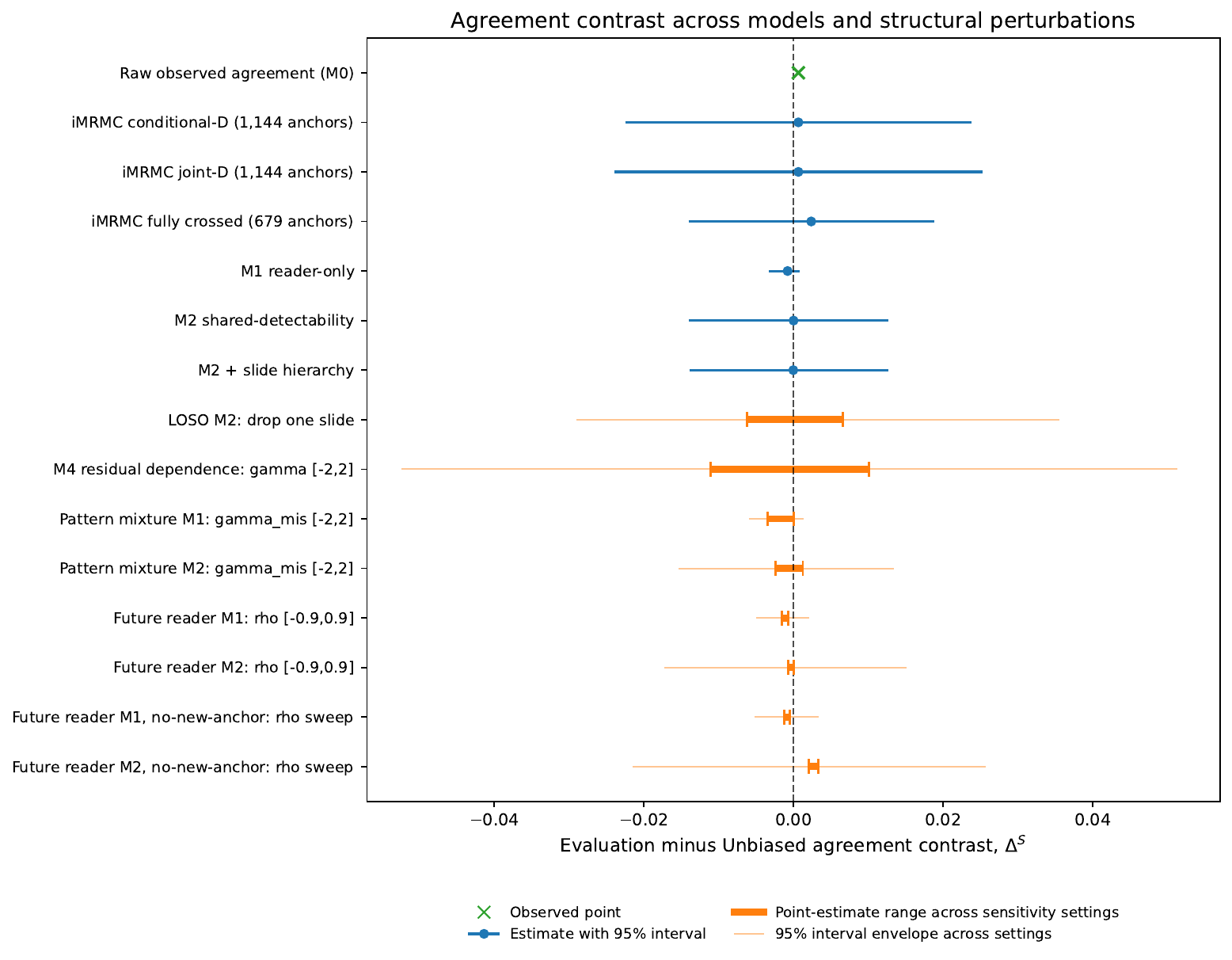}
\caption{\label{fig:delta}Evaluation-minus-Unbiased fixed-AI/future-reader agreement contrast across empirical benchmarks, Bayesian models, frames, and structural sensitivities. For fitted models and MRMC benchmarks, points and intervals show the corresponding inferential estimates. For sensitivity rows, thick segments show the range of point estimates across prespecified settings and thin segments show the envelope of associated 95\% intervals. These ranges are neither posterior intervals over the sensitivity parameter nor formal partial-identification bounds.}
\end{figure}

\subsection{Structural sensitivity of absolute agreement}

Absolute agreement was much more sensitive than the condition contrast. M1 placed future-reader agreement near 0.528 in both conditions, whereas M2 placed it near 0.579. The slide-hierarchy extension left the full-data M2 values essentially unchanged, but the five LOSO refits produced point-estimate ranges of 0.4990 to 0.6464 for Unbiased and 0.5056 to 0.6402 for Evaluation. The corresponding interval envelopes extended from 0.4823 to 0.6593 and from 0.4759 to 0.6582, respectively.

The strongest structural movement came from M3. Across $\gamma\in[-2,2]$, the Unbiased point estimate ranged from 0.4205 to 0.6292 and the Evaluation estimate from 0.4094 to 0.6393. The 95\% interval envelopes extended to 0.4003--0.6503 for Unbiased and 0.3640--0.6843 for Evaluation. In contrast, the pattern-mixture missingness grid moved M2 by less than about 1.1 percentage points in either condition, and the $\rho_{ub}$ future-reader sensitivity moved the absolute levels by roughly a tenth of a percentage point or less under the full construction.

\begin{figure}[htbp]
\centering
\includegraphics[width=0.8\textwidth]{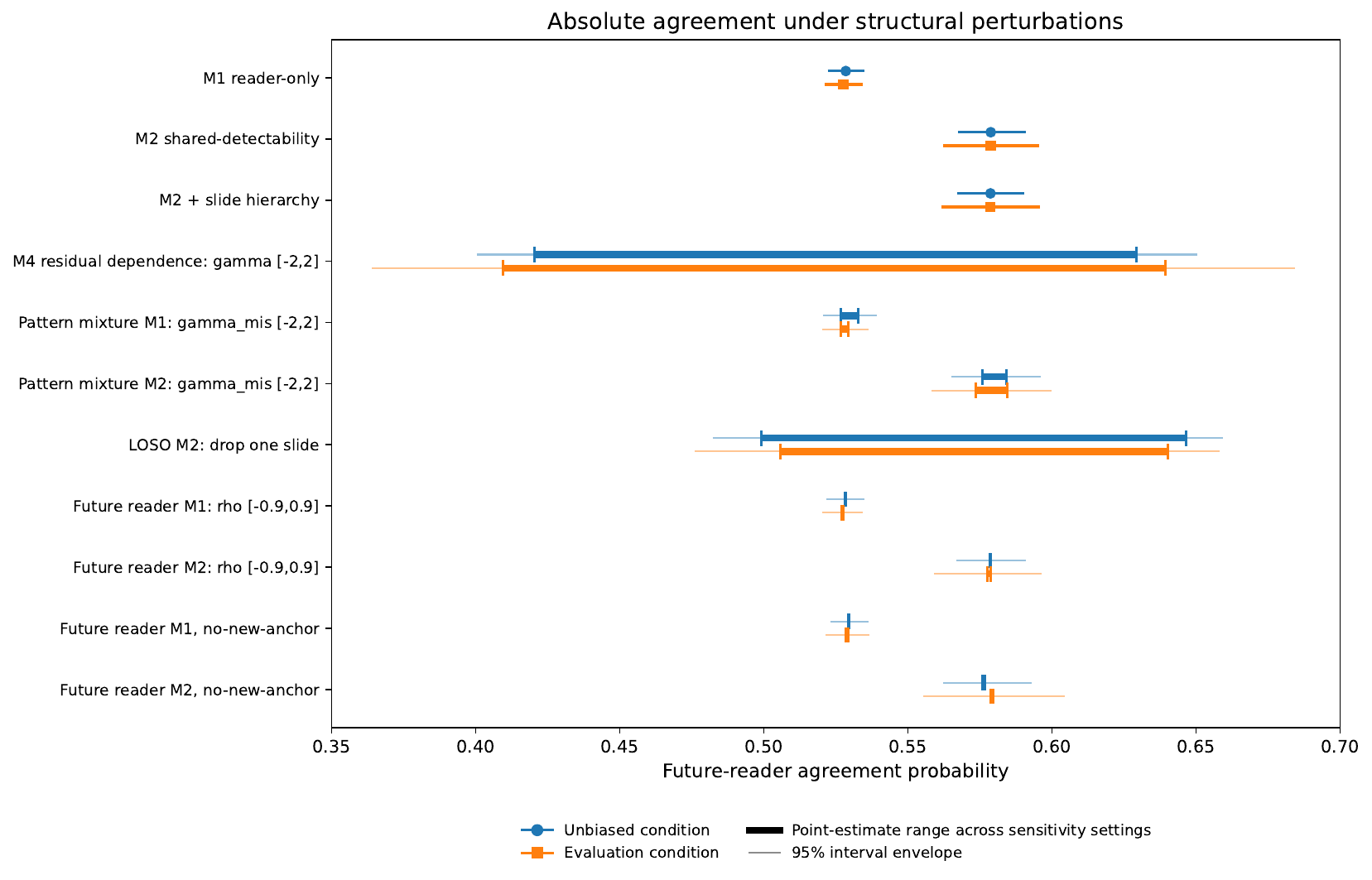}
\caption{\label{fig:absolute}Structural sensitivity of condition-specific fixed-AI/future-reader agreement. Point estimates and 95\% credible intervals are shown for fitted models; sensitivity rows summarize the movement of point estimates and the envelope of intervals across prespecified settings. LOSO ranges characterize sensitivity to the composition of the five observed slides/patients and are not estimates of new-patient transportability. Shared detectability is a working dependence structure, and no displayed structural range is interpreted as a formal identification bound.}
\end{figure}

Thus, across the principal prespecified structural point-estimate ranges, absolute agreement extended from approximately 0.420 to 0.646 in Unbiased and from 0.409 to 0.640 in Evaluation. Figure~\ref{fig:absolute} shows these condition-specific estimates on the same agreement scale. The separation between M1 and M2 illustrates sensitivity to whether AI and human responses share a working detectability structure; the LOSO rows show sensitivity to which slide is omitted; and the M3 rows show the larger movement allowed by residual AI--human dependence. In contrast, the missingness and reader-correlation perturbations produce comparatively narrow ranges. The figure therefore emphasizes the full structural range rather than privileging one model as the uniquely correct decomposition.

\subsection{Model adequacy, simulation, and predictive evaluation}

\subsubsection{Prior sensitivity and parameter dependence}

Across the examined weaker and more-regularizing M2 prior families, the range of absolute agreement point estimates was about 0.54 percentage points and the range of $\Delta^S$ was about 0.093 percentage points, substantially smaller than the M1--M2 structural shift in the absolute levels. This supports prior robustness of the derived estimands over the examined prior range, but it does not establish unique latent-factor identification. Detailed prior-family estimates are reported in Supplementary Table S2, and selected posterior parameter correlations are reported in Supplementary Table S3.

Posterior dependence diagnostics made that distinction explicit. The strongest pairwise dependence among the requested M2 quantities was $\mathrm{corr}(\lambda_A,\sigma_a)=-0.682$, followed by $\mathrm{corr}(\sigma_f,\sigma_a)=0.505$ and $\mathrm{corr}(\lambda_A,\sigma_f)=-0.456$. These tradeoffs indicate that related portions of the observed variation can be allocated differently among the shared latent components. They are not proof of formal nonidentifiability, but they argue against treating the exact variance decomposition as uniquely verified.

The explicit slide-hierarchy fit was computationally clean and estimated a slide-scale posterior mean of 0.638, median 0.607, with an equal-tail 95\% interval of 0.354 to 1.099. With only five slides/patients this variance component is not estimated precisely as a population quantity. Its main value is diagnostic: the full-data hierarchy barely changed the selected-frame estimands, whereas the LOSO refits showed large composition sensitivity in the absolute levels.

\subsubsection{Performance under structural misspecification}

Table~\ref{tab:adversarial} summarizes the frozen A1--A8 stress program. The prespecified comparative-stability criterion passed: median $R$ across the 16 M1/M2 cells was 0.295 and 14/16 cells had $R<1$. The only $R\ge1$ cells occurred in A4, which imposed condition-specific AI--human dependence. In A4, M1 and M2 had $\Delta^S$ biases of approximately $-2.48$ and $-2.27$ percentage points, respectively, and both had 0\% interval coverage for the contrast. The full M3 sensitivity envelope contained the true A4 contrast in 94\% of replications, compared with 100\% in A1--A3 and A5--A8. Thus, condition-specific dependence is a clear robustness boundary rather than a hidden exception. A small $R$ should not be mistaken for globally good model calibration: for example, M1 had low $R$ in several scenarios where coverage of the absolute agreement levels was poor. The ratio was prespecified specifically to compare contrast recovery with absolute-level recovery, not to rank overall method quality.

Supplementary Table S7 documents the data-generating departures and frozen truth values, Supplementary Table S8 reports absolute-target coverage with Monte Carlo standard errors, and Supplementary Table S9 reports M3 sensitivity-envelope containment.

\begin{table}[H]
\caption{Adversarial misspecification results} 
\label{tab:adversarial}
\centering
\begin{threeparttable}
\footnotesize
\setlength{\tabcolsep}{3.5pt}
\begin{tabular}{llrrrrrrrrrr}
\toprule
Scenario & Model & $n$ & $\theta_U$ RMSE & $\theta_U$ cov. & $\theta_E$ RMSE & $\theta_E$ cov. & $\Delta$ bias & $\Delta$ RMSE & $\Delta$ cov. & $R$ & M3 contain. \\
\midrule
A1 & M1 & 100 & 0.235 & 98\% & 0.254 & 98\% & 0.005 & 0.069 & 99\% & 0.282 & 100\% \\
A1 & M2 & 99 & 0.462 & 82.8\% & 0.492 & 83.8\% & 0.025 & 0.080 & 97.0\% & 0.167 & 100\% \\
A2 & M1 & 100 & 1.283 & 3\% & 1.365 & 4\% & -0.078 & 0.117 & 75\% & 0.088 & 100\% \\
A2 & M2 & 100 & 0.491 & 93\% & 0.523 & 96\% & 0.004 & 0.156 & 99\% & 0.309 & 100\% \\
A3 & M1 & 100 & 2.323 & 0\% & 2.517 & 0\% & -0.193 & 0.204 & 32\% & 0.084 & 100\% \\
A3 & M2 & 100 & 0.369 & 99\% & 0.395 & 96\% & -0.066 & 0.170 & 98\% & 0.446 & 100\% \\
A4 & M1 & 100 & 0.633 & 63\% & 3.062 & 0\% & -2.484 & 2.496 & 0\% & 1.351 & 94\% \\
A4 & M2 & 100 & 1.396 & 21\% & 1.085 & 67\% & -2.271 & 2.288 & 0\% & 1.845 & 94\% \\
A5 & M1 & 100 & 2.178 & 16\% & 2.356 & 21\% & -0.169 & 0.321 & 80\% & 0.142 & 100\% \\
A5 & M2 & 100 & 0.498 & 99\% & 0.575 & 99\% & -0.058 & 0.350 & 99\% & 0.653 & 100\% \\
A6 & M1 & 100 & 2.237 & 1\% & 2.396 & 2\% & -0.156 & 0.201 & 72\% & 0.087 & 100\% \\
A6 & M2 & 100 & 0.505 & 99\% & 0.556 & 98\% & -0.014 & 0.230 & 96\% & 0.434 & 100\% \\
A7 & M1 & 100 & 2.415 & 0\% & 2.580 & 0\% & -0.162 & 0.270 & 95\% & 0.108 & 100\% \\
A7 & M2 & 100 & 0.421 & 98\% & 0.451 & 99\% & -0.052 & 0.137 & 99\% & 0.314 & 100\% \\
A8 & M1 & 100 & 2.461 & 0\% & 2.623 & 0\% & -0.163 & 0.196 & 70\% & 0.077 & 100\% \\
A8 & M2 & 100 & 0.548 & 95\% & 0.549 & 98\% & -0.022 & 0.210 & 99\% & 0.384 & 100\% \\
\bottomrule
\end{tabular}
\begin{tablenotes}[flushleft]
\scriptsize 
\item \textit{Note.} RMSE and bias are reported in percentage points, and coverage is empirical 95\% interval coverage. The comparative-stability ratio \(R\) is defined in Equation~\eqref{eq:R}. The M3 containment column reports whether the full M3 sensitivity envelope contained the true \(\Delta^S\). A1 represents independent AI-specific item variation; A2, correlated but nonidentical AI and human factors; A3, a nonlinear or threshold AI response; A4, condition-specific AI--human dependence; A5, an omitted slide or patient effect; A6, informative FOV missingness; A7, selection on AI and human positivity; and A8, mixture or heavy-tailed heterogeneity. M2 in A1 is based on 99 valid replications because one prespecified replication did not satisfy the computational gate after targeted retry. The table identifies robustness boundaries and should not be interpreted as an overall ranking of the methods. \end{tablenotes}
\end{threeparttable}
\end{table}

\subsubsection{Predictive assessment}

All 15 Bayesian LOSO predictive fits passed the computational gate (0 divergences; maximum scalar $\widehat R=1.00$; minimum bulk ESS 1,498; minimum tail ESS 1,732). In pooled anchor-level cross-validation, held-out-AI prevalence was 0.518. The latent Bayesian predictor achieved Brier score 0.1846, log loss 0.5463, AUROC 0.7957, and AUPRC 0.8155. Nested regularized logistic regression yielded 0.2292, 0.6477, 0.6472, and 0.6237, respectively; histogram gradient boosting yielded 0.2302, 0.6523, 0.6429, and 0.6044. These results support shared predictive information between human observations and the fixed AI within the selected frame; they do not establish AI correctness or uniquely validate M2. Held-out-AI calibration intercept/slope were $-0.033/1.121$ for the latent Bayesian predictor, 0.005/0.973 for regularized logistic regression, and 0.026/0.739 for gradient boosting, so the predictive comparison also differed in calibration shape rather than only discrimination. Condition-specific predictive metrics and reliability curves are reported in Supplementary Table S11 and Supplementary Figure S2; nested tuning and leave-one-slide-out diagnostics are reported in Supplementary Tables S10 and S12 and Supplementary Figure S3.

Held-out-human differences between joint and reader-only predictions were small. Their pooled Brier scores were 0.1329 and 0.1336 and log losses 0.4301 and 0.4320, respectively. For target-aligned agreement, observed prevalence was 0.5469. Joint and reader-only AUROCs were 0.8736 and 0.8741 and AUPRC values were 0.8783 and 0.8778, showing essentially unchanged discrimination. Their calibration-in-the-large differed: mean predicted agreement was 0.5598 for the joint model and 0.5161 for the reader-only model. For target-aligned agreement, calibration intercept/slope were $-0.095/0.977$ for the joint model and $0.233/1.001$ for the reader-only model, highlighting that similar discrimination can coexist with different calibration.

LOSO results were much less stable. Held-out-AI Brier scores ranged from 0.2248 to 0.4271, AUROC from 0.2988 to 0.6516, and AUPRC from 0.1798 to 0.8667 as the held-out slide changed. For target-aligned agreement, the joint model had worse Brier and log-loss performance on four of five held-out slides and slightly better performance on one. These findings reinforce the distinction between within-context shared information and slide-composition robustness.

Table~\ref{tab:predictive} consolidates the pooled predictive results. Panel A reports proper scoring rules and discrimination for the three prediction targets. For held-out AI, the latent Bayesian predictor had the lowest Brier score and log loss and the highest AUROC and AUPRC among the three compared predictors. For held-out human responses and target-aligned agreement, joint and reader-only models performed similarly in discrimination and proper scoring. Panel B shows that this similarity did not imply identical calibration: the joint and reader-only target-aligned models had comparable slopes but different calibration intercepts. The bracketed slide-bootstrap intervals are broad in several cells and should be read as evidence of slide-composition sensitivity rather than precise patient-population uncertainty.

\begin{figure}[H]
\centering
\includegraphics[width=0.85\textwidth]{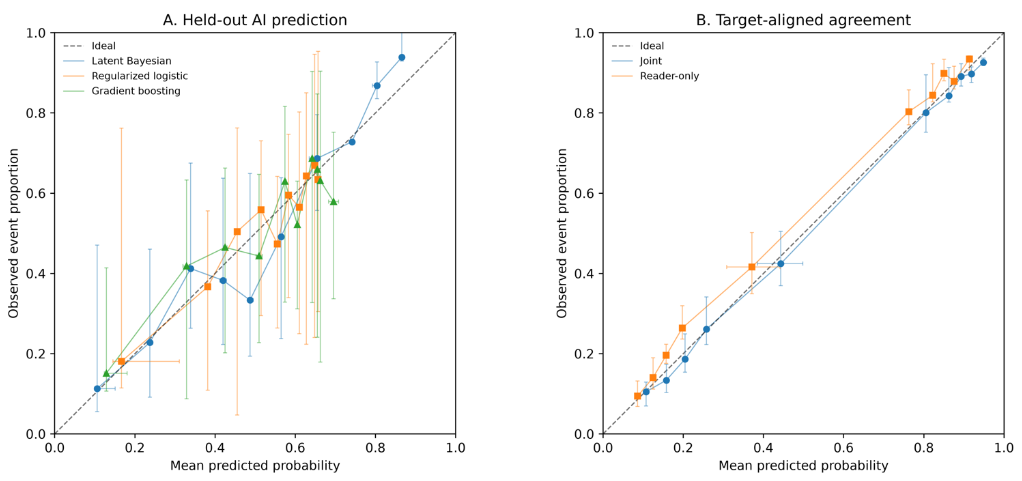}
\caption{\label{fig:predictive}Predictive reliability for held-out AI decisions and target-aligned agreement on the selected frame. Panel A shows held-out-AI probabilities from the latent Bayesian predictor, nested regularized logistic regression, and histogram gradient boosting. Human observations from the held-out anchor remain available to the latent predictor, so this tests shared information rather than AI correctness. Panel B shows target-aligned agreement probabilities under joint and reader-only human-response models. Error bars are slide-cluster bootstrap summaries; with five slides, they represent composition sensitivity rather than precise patient-population uncertainty. The outer folds hold out anchors while retaining all FOVs and slides and therefore do not constitute new-patient validation.}
\end{figure}

\begin{table}[H]
\caption{Predictive performance and calibration} 
\label{tab:predictive}
\centering
\begin{threeparttable}
\footnotesize
\setlength{\tabcolsep}{3.5pt}
\noindent\parbox{\textwidth}{\raggedright\textbf{Panel A. Proper scoring and discrimination}}\\
\begin{tabularx}{\textwidth}{p{0.17\textwidth}p{0.17\textwidth}rrXXXX}
\toprule
Target & Model & $N$ & Prev. & Brier & Log loss & AUROC & AUPRC \\
\midrule
Held-out AI & Latent Bayesian & 1,144 & 0.518 & 0.185 [0.163,0.212] & 0.546 [0.495,0.612] & 0.796 [0.696,0.824] & 0.816 [0.517,0.870] \\
Held-out AI & Regularized logistic & 1,144 & 0.518 & 0.229 [0.218,0.247] & 0.648 [0.626,0.683] & 0.647 [0.578,0.671] & 0.624 [0.286,0.838] \\
Held-out AI & Gradient boosting & 1,144 & 0.518 & 0.230 [0.217,0.248] & 0.652 [0.625,0.689] & 0.643 [0.593,0.663] & 0.604 [0.326,0.798] \\
Held-out human & Joint & 10,833 & 0.834 & 0.133 [0.113,0.144] & 0.430 [0.379,0.458] & 0.653 [0.633,0.674] & 0.896 [0.876,0.917] \\
Held-out human & Reader-only & 10,833 & 0.834 & 0.134 [0.114,0.145] & 0.432 [0.378,0.462] & 0.644 [0.615,0.687] & 0.894 [0.870,0.922] \\
Target-aligned agreement & Joint & 10,833 & 0.547 & 0.133 [0.110,0.144] & 0.430 [0.372,0.460] & 0.874 [0.832,0.904] & 0.878 [0.792,0.911] \\
Target-aligned agreement & Reader-only & 10,833 & 0.547 & 0.134 [0.111,0.145] & 0.432 [0.372,0.461] & 0.874 [0.827,0.898] & 0.878 [0.778,0.911] \\
\bottomrule
\end{tabularx}

\vspace{2mm}
\textbf{Panel B. Calibration}
\begin{tabularx}{\textwidth}{p{0.27\textwidth}p{0.25\textwidth}XX}
\toprule
Target & Model & Calibration intercept & Calibration slope \\
\midrule
Held-out AI & Latent Bayesian & -0.033 [-0.782,0.644] & 1.121 [0.724,1.261] \\
Held-out AI & Regularized logistic & 0.005 [-1.196,0.849] & 0.973 [0.548,1.213] \\
Held-out AI & Gradient boosting & 0.026 [-1.066,0.877] & 0.739 [0.475,0.863] \\
Held-out human & Joint & 0.248 [0.095,0.335] & 0.834 [0.777,0.984] \\
Held-out human & Reader-only & -0.038 [-0.398,0.147] & 1.019 [0.875,1.292] \\
Target-aligned agreement & Joint & -0.095 [-0.165,0.014] & 0.977 [0.901,1.079] \\
Target-aligned agreement & Reader-only & 0.233 [0.185,0.333] & 1.001 [0.922,1.125] \\
\bottomrule
\end{tabularx}
\begin{tablenotes}[flushleft]
\scriptsize
\item \textit{Note.} Panel A summarizes pooled nested anchor-level predictive performance for held-out AI decisions, held-out human responses, and target-aligned agreement. Panel B reports calibration intercepts and slopes. Bracketed values are 95\% slide-cluster bootstrap intervals. AUPRC should be interpreted relative to the observed prevalence of the corresponding prediction target. The held-out-anchor analyses assess prediction within the observed slide and FOV contexts and do not constitute new-patient or external validation.
\end{tablenotes}
\end{threeparttable}
\end{table}

Figure~\ref{fig:predictive} complements Table~\ref{tab:predictive} by displaying reliability rather than only aggregate scores. In Panel A, predicted held-out-AI probabilities are compared with observed frequencies for the latent Bayesian, regularized logistic, and gradient-boosting predictors. In Panel B, the same comparison is shown for target-aligned agreement under the joint and reader-only models. Curves closer to the diagonal indicate better agreement between predicted probabilities and observed frequencies, while the slide-cluster error bars show how strongly those frequencies vary with the small set of slides. The figure supports shared predictive information within the observed context, but the uncertainty and slide-to-slide variation do not support new-patient transportability or unique validation of the M2 latent structure.

\section{Discussion}

Four findings organize the strengthened analysis. First, the average Evaluation-minus-Unbiased fixed-AI agreement contrast remained near zero across raw agreement, iMRMC, M1, M2, the slide hierarchy, and several structural sensitivities in the selected frame. Second, that near-zero average did not imply stable individual annotations: approximately one in five paired reader-anchor observations changed AI-agreement state, with nearly equal numbers switching in opposite directions. Third, the absolute level of future-reader agreement changed materially when the reference-process model or slide composition changed. Fourth, the contrast and absolute levels therefore require different reporting: a comparatively stable descriptive contrast can coexist with structurally fragile absolute agreement.

The cancellation revealed by Figure~\ref{fig:raw} is central. Marginal agreement was almost unchanged between workflows, yet 19.62\% of paired observations moved between agreement states. A near-zero average contrast can therefore arise from opposing reader-level changes rather than from invariant labels. This distinction is important for human--AI evaluation because a stable average metric may conceal different underlying error or disagreement patterns.

Absolute agreement was less stable. The approximately five-percentage-point shift from M1 to M2, the 13--15 percentage-point LOSO ranges, and the much wider M3 sensitivity ranges all show that one absolute agreement number can depend strongly on how the human reference process and AI-human residual association are represented. This is precisely the type of uncertainty that is hidden if multiple readers are collapsed into one consensus reference and the resulting label is treated as fixed. The broader literature on expert-panel standards, pathology annotation variability, and multi-annotator learning reaches the same methodological warning from different directions: the reference process must be characterized rather than assumed away.\cite{Kellerhuis2025,Wahab2022,Zhang2023,Elmore2015} The findings are consistent with current medical-imaging guidance that reference standards and observer variability should be documented explicitly\cite{Tejani2024,Hadjiiski2023,Hanna2024} and with recent pathology evidence that measured AI performance can depend on which experts define the reference panel.\cite{Shamshoian2026}

The predictive results support shared information but not a unique latent decomposition. Human-informed latent structure was associated with held-out AI decisions under the observed-context anchor-level design, whereas incremental gains for held-out human labels were small and target-aligned discrimination was essentially unchanged between joint and reader-only predictions. The latent-factor interpretation is further tempered by posterior tradeoffs between $\lambda_A$ and shared heterogeneity scales. Thus, the data are consistent with common AI-human difficulty structure, but they do not establish that M2 is the uniquely correct factorization or that a positive AI loading has biological meaning.

The adversarial program makes the robustness boundary more explicit. Under seven of eight prespecified departures, $\Delta^S$ was comparatively more stable than the two absolute agreement levels for both M1 and M2. A4 deliberately violated the cancellation structure by making AI-human dependence condition specific, and it defeated both Bayesian models: contrast bias was about 2.3--2.5 percentage points and coverage was 0\%. M3 reduced sensitivity to residual dependence but did not universally absorb this condition-specific failure. The appropriate conclusion is therefore not that the Bayesian models are robust in general; it is that the contrast was comparatively stable under many examined perturbations, with a named mechanism under which that stability failed.

These findings complement rather than replace MRMC methods. Binary MRMC remains an appropriate nonlatent benchmark when the scientific target is an observed reader/case score and the reader/case sampling design supports population inference.\cite{Chen2014,Obuchowski2022} The present framework is useful when the reference-reader process itself is part of the uncertainty and the investigator wants to distinguish what is directly observed, what is generalized to a future reader, and what changes under alternative reference-process assumptions. Likewise, latent-class or consensus models remain useful when an unobserved disease or consensus state is the scientific target.\cite{DawidSkene1979,Warfield2004,CollinsHuynh2014} The contribution here is to keep those targets distinct rather than ranking one family universally above another.

For developers and medical-imaging investigators, a practical reporting implication follows. When human labels are variable, an AI evaluation should state at least: the reader population being generalized to; whether cases/patients are fixed or sampled; how nonannotation differs from a negative label; whether AI influenced the labeling workflow; whether the AI being evaluated is the same system used during annotation; the selected-frame definition; and how agreement conclusions change under plausible reference-process and missingness assumptions. For regulators and reader-study designers, a single agreement percentage is insufficient when the uncertainty in the reference process is large relative to conventional sampling uncertainty.

\subsection{Implications for medical-imaging AI studies}

For model evaluation and reader-study reporting, the results support a concise minimum record: identify the readers and annotation workflow; distinguish missing observations from observed negative decisions; state whether inference targets the observed readers or a future reader population; report directional switching as well as marginal agreement; separate sampling uncertainty from alternative reference-process assumptions; describe the case, slide, or patient composition to which the estimate applies; and avoid interpreting agreement as accuracy without an independent reference designed for that purpose. These elements allow investigators to see whether a stable average result reflects stable annotations or offsetting changes and whether the reported absolute value is robust to plausible reference-process choices.

\subsection{Limitations}

The first limitation concerns interpretation. The comparison between Evaluation and Unbiased annotations is descriptive because workflow condition was not randomized independently of study order. Agreement with the fixed AI is also not diagnostic accuracy or biological correctness: a reader and an AI can agree while both are wrong, and opposing annotation changes can leave average agreement nearly unchanged. The results therefore characterize agreement under the observed workflows and selected frame, not a causal assistance effect or equivalence between workflows.

The second limitation concerns generalizability. The reader-population distribution was estimated from five primary readers, and the selected frame contained five slides/patients and 52 FOVs. The future-reader estimand generalizes only to qualified readers plausibly exchangeable with these participants and remains conditional on the selected anchors, slides, and frozen HoVer-Net decisions. The slide-hierarchy and leave-one-slide-out analyses quantify sensitivity to the composition of these five slides; they do not constitute independent new-patient or external validation. The anchor-held-out predictive analyses similarly evaluate shared information within the observed slide/FOV structure rather than transport to a new institution or algorithm class.

The third limitation is structural. Absolute agreement depends on how reader heterogeneity, shared detectability, residual AI--human association, and missing observations are represented. The sensitivity analyses describe prespecified perturbations rather than probabilities over all possible data-generating mechanisms or sharp identification bounds. In particular, condition-specific AI--human dependence was the common boundary at which contrast stability failed in simulation. These findings motivate reporting a range of structurally defensible estimates rather than treating one fitted model as uniquely correct.

\subsection{Future work}

Replication should include more readers, slides, institutions, and frozen AI systems so that reader-, patient-, and algorithm-level transportability can be examined directly. Studies targeting causal effects of assistance would require randomized reader--FOV assignment and protection against order or memory effects. Studies targeting biological accuracy would require an independent blinded reference process designed for that purpose.

\section{Conclusion}

In this selected NuCLS frame, the average difference in fixed-AI agreement between Evaluation and Unbiased annotations remained near zero across direct, MRMC, and Bayesian analyses, even though approximately one in five paired annotations changed agreement state. Absolute agreement with a future reader was more sensitive to the assumed reference process and to slide composition. The practical implication is straightforward: medical-imaging AI studies should report who produced the reference labels, how missing and discordant annotations were handled, and how conclusions change under defensible alternative reference-process assumptions. A single agreement estimate and its sampling interval do not fully describe uncertainty when the reference labels themselves vary.

\subsection*{Disclosures}
The author declares no competing interests.

\subsection*{Code, Data, and Materials Availability}

The NuCLS source data are publicly available under the license and access terms reported by Amgad et al.\cite{Amgad2022} The source dataset is not redistributed in the study repository; users must obtain it under the original access and license terms. The study used a fixed external HoVer-Net checkpoint with SHA-256 hash \url{cc6870b201592ec4715afa558ef781a0e59fa4ebc4fb4d974d3c8319ce379f2b}. Analysis code, dataset-access instructions, preprocessing scripts, frozen simulation materials, derived result tables, figure-generation scripts, and environment specifications are available at \url{https://github.com/haolayinka/Nucls-ai-uq}.

\subsection*{Acknowledgments}
Results in this paper were obtained using a high-performance computing system acquired through NSF MRI grant DMS-1337943 to Worcester Polytechnic Institute. During manuscript preparation, the author used OpenAI ChatGPT and Microsoft 365 Copilot to assist with language and grammar cleanup, consistency checks, LaTeX preparation, and reference formatting. These tools were not used to generate data, conduct the statistical analyses, determine the scientific conclusions, or assume authorship responsibility. The author reviewed and edited all AI-assisted content, verified the numerical results and scientific interpretations against the frozen analysis outputs, and takes full responsibility for the content of the manuscript.


\bibliography{report}
\bibliographystyle{spiejour}

\end{spacing}
\end{document}


\begin{center}
{\Large\bfseries Supplementary Material\par}
\vspace{0.6em}
{\large Bayesian Uncertainty Quantification of Breast-Cancer Pathology AI Agreement under Reader- and Annotation-Condition-Dependent Reference Labels\par}
\end{center}
\vspace{1em}

\setcounter{table}{0}
\renewcommand{\thetable}{S\arabic{table}}
\setcounter{figure}{0}
\renewcommand{\thefigure}{S\arabic{figure}}
\setcounter{equation}{0}
\renewcommand{\theequation}{S\arabic{equation}}

\section{Purpose and scope}
This supplement provides mathematical derivations, prior and identification analyses, structural-sensitivity details, simulation mechanisms, and predictive diagnostics that are not reproduced in the main article. Primary selected-frame summaries, principal point estimates, and the main adversarial and predictive tables are reported only in the article.

\section{Notation bridge} 
The main article defines M1, M2, the baseline priors, and the selected-frame estimands. The sections below provide only the additional derivations, construction sensitivities, diagnostic details, and simulation mechanisms needed to reproduce or interrogate those analyses.

\section{Derivation of the future-reader agreement estimand}
\subsection{Posterior predictive probability}
Let $c\in\{U,E\}$ index annotation condition, $i$ index anchor, $f(i)$ denote its field of view, and $a_i\in\{0,1\}$ be the fixed AI decision. Conditional on a posterior draw $\psi$ and a future reader effect vector $R^\star$, write the human linear predictor as
\[
\eta_{ic}(R^\star,\psi)=m_{ic}(\psi)+z_c^{\mathsf T}R^\star,
\]
where $m_{ic}(\psi)$ contains the fitted fixed effects and the anchor/FOV latent components retained in the selected-frame estimand. The future-reader detection probability is
\[
p_{ic}(\psi)=\int \operatorname{logit}^{-1}\!\left\{\eta_{ic}(r,\psi)\right\}
\phi_q(r;0,\Sigma_R)\,dr.
\]
For M1, $m_{ic}$ is obtained from the human-response model without an AI likelihood. For M2, $m_{ic}$ contains the shared detectability components inferred jointly from the human and fixed-AI channels.

\subsection{Logistic-normal reduction}
Because $z_c^{\mathsf T}R^\star$ is Gaussian, define
\[
v_c=z_c^{\mathsf T}\Sigma_R z_c,
\qquad Z\sim N(0,1).
\]
Then the $q$-dimensional integral reduces exactly to
\[
p_{ic}(\psi)=\int \operatorname{logit}^{-1}\!\left(m_{ic}(\psi)+\sqrt{v_c}\,z\right)\phi(z)\,dz.
\]
With $K$ Gauss--Hermite nodes $x_k$ and weights $w_k$ under the convention
$\int e^{-x^2}g(x)dx\approx\sum_{k=1}^K w_kg(x_k)$,
\[
p_{ic}(\psi)\approx \frac{1}{\sqrt{\pi}}
\sum_{k=1}^{K} w_k\,
\operatorname{logit}^{-1}\!\left(m_{ic}(\psi)+\sqrt{2v_c}\,x_k\right).
\]

\subsection{Agreement probability and selected-frame averaging}
Since $a_i$ is fixed,
\[
q_{ic}(\psi)=\Pr(Y_{ic}^{\star}=a_i\mid\psi)
=a_i p_{ic}(\psi)+(1-a_i)\{1-p_{ic}(\psi)\}.
\]
For $N$ selected anchors,
\[
\theta_c^S(\psi)=\frac{1}{N}\sum_{i=1}^{N}q_{ic}(\psi),
\qquad
\Delta^S(\psi)=\theta_E^S(\psi)-\theta_U^S(\psi).
\]
Posterior summaries are obtained by evaluating these functions draw by draw. Thus posterior uncertainty in model parameters and future-reader variation is propagated jointly, while the selected anchors, FOVs, slides, and AI vector remain fixed by definition.

\subsection{Correlated future-reader effects}
If the future-reader intercept $u^\star$ and condition effect $b^\star$ have standard deviations $\sigma_u,\sigma_b$ and correlation $\rho_{ub}$,
\[
\binom{u^\star}{b^\star}\sim N\!\left[
\binom00,
\begin{pmatrix}
\sigma_u^2 & \rho_{ub}\sigma_u\sigma_b\\
\rho_{ub}\sigma_u\sigma_b & \sigma_b^2
\end{pmatrix}\right].
\]
Consequently,
\[
v_U=\sigma_u^2,
\qquad
v_E=\sigma_u^2+\sigma_b^2+2\rho_{ub}\sigma_u\sigma_b,
\]
with additional reader-by-FOV and reader-by-anchor variances added to both expressions where present. This identity is the basis of the fixed-$\rho_{ub}$ construction sensitivity.

\section{Likelihood factorization and identification constraints}
Let $O_{ric}$ indicate that reader $r$ supplied an observed response for anchor $i$ under condition $c$. The observed human likelihood is
\[
L_H(\psi)=\prod_{r,i,c:O_{ric}=1}
\pi_{ric}^{y_{ric}}(1-\pi_{ric})^{1-y_{ric}}.
\]
M1 uses $L_H(\psi)$ only. M2 augments it with one AI observation per anchor,
\[
L_{M2}(\psi)=L_H(\psi)\prod_i
\pi_{Ai}^{a_i}(1-\pi_{Ai})^{1-a_i},
\quad
\operatorname{logit}(\pi_{Ai})=\alpha_A+\lambda_A d_i.
\]
The shared latent scale is identified by its Gaussian variance specification and the orientation constraint $\lambda_A>0$, implemented through a half-normal prior. These conventions identify a working parameterization; they do not establish a unique biological latent decomposition.

\section{Future-reader integration and construction sensitivity}

Under the baseline independent reader-effect construction,
\[
\sigma_U^2=\sigma_r^2+\sigma_{rf}^2+\sigma_{ra}^2,
\]
\[
\sigma_E^2=\sigma_r^2+\sigma_{rE}^2+\sigma_{rf}^2+\sigma_{ra}^2.
\]

For posterior draw \(m\), future-reader response probabilities are logistic-normal expectations over a new future reader. Agreement with the frozen AI is
\[
q_{ic}^{(m)}
=A_i p_{ic}^{*(m)}
+(1-A_i)\{1-p_{ic}^{*(m)}\},
\]
and \(\theta_c^S\) is the selected-anchor average of \(q_{ic}^{(m)}\).

A fresh reader-by-anchor draw on an already observed anchor is interpreted as reader-specific perception/interpretation of that same nucleus, not as a new biological state. Because this choice is scientific rather than directly identified, an alternate construction omits new reader-by-anchor variation from future-reader integration while retaining it in the observed-reader likelihood.

Correlation sensitivity fixes
\[
\rho_{ub}\in\{-0.90,-0.75,-0.50,-0.25,0,0.25,0.50,0.75,0.90\}.
\]
In Evaluation, the marginal reader variance includes \(2\rho_{ub}\sigma_r\sigma_{rE}\).

Table~\ref{tab:S-reader-correlation} summarizes the sensitivity of the selected-frame estimands to the fixed future-reader correlation grid and to inclusion or exclusion of new reader-by-anchor variation.
\begin{table}[H]
\centering
\setlength{\tabcolsep}{12pt}
\begin{threeparttable}
\caption{Future-reader correlation sensitivity}
\label{tab:S-reader-correlation}
\begin{tabular}{lrrr}
\toprule
Model / future-reader construction & $\theta_U$ span & $\theta_E$ span & $\Delta$ span\\
\midrule
M1 full & 0.0392 & 0.0771 & 0.0843\\
M1 no-new-reader-anchor & 0.0329 & 0.0761 & 0.0784\\
M2 full & 0.0391 & 0.1056 & 0.0760\\
M2 no-new-reader-anchor & 0.0711 & 0.0966 & 0.1229\\
\bottomrule
\end{tabular}
\begin{tablenotes}[flushleft]
\scriptsize
\item \textit{Note.} Values are from the frozen, adjudicated analysis outputs. Spans are reported on the probability scale unless explicitly identified as percentage points. M1 is the reader-only model; M2 is the shared-detectability model; M3 is the residual-dependence sensitivity.
\end{tablenotes}
\end{threeparttable}
\end{table}

For M2, the no-new-reader-anchor construction shifts \(\Delta\) relative to the full construction by approximately +0.238 to +0.386 percentage points over the \(\rho\) grid; at \(\rho=0\), the shift is approximately +0.278 percentage points. The construction choice is more consequential than the bounded reader-effect correlation in this dataset.

\section{Prior sensitivity and identification diagnostics}

\subsection{Derived estimand sensitivity}

Table~\ref{tab:S-prior-sensitivity} reports the derived estimands under the four prespecified alternative prior families.
\begin{table}[H]
\centering
\setlength{\tabcolsep}{12pt}
\begin{threeparttable}
\caption{Prior sensitivity}
\label{tab:S-prior-sensitivity}
\begin{tabular}{lrrr}
\toprule
Prior family & $\theta_U$ & $\theta_E$ & $\Delta$\\
\midrule
More-regularizing AI-loading prior & 0.581442 & 0.581393 & -0.000050\\
Weaker AI-loading prior & 0.576062 & 0.576109 & +0.000047\\
More-regularizing variance priors & 0.577520 & 0.578034 & +0.000514\\
Weaker variance priors & 0.579027 & 0.578615 & -0.000412\\
\bottomrule
\end{tabular}
\begin{tablenotes}[flushleft]
\scriptsize
\item \textit{Note.} Values are from the frozen, adjudicated analysis outputs. Spans are reported on the probability scale unless explicitly identified as percentage points. M1 is the reader-only model; M2 is the shared-detectability model; M3 is the residual-dependence sensitivity.
\end{tablenotes}
\end{threeparttable}
\end{table}

The absolute estimates move by roughly 0.54 percentage points across the examined prior families, while \(\Delta\) moves by roughly 0.093 percentage points. This is prior robustness of the derived estimands over the examined range, not proof that latent components are uniquely identified.

\subsection{Posterior parameter tradeoffs}

Diagnostics were computed from 16,000 M2 posterior draws.

Table~\ref{tab:S-identification-correlations} reports pairwise posterior correlations among the principal latent-loading and variance parameters.
\begin{table}[H]
\centering
\setlength{\tabcolsep}{26pt}
\begin{threeparttable}
\caption{Posterior identification correlations}
\label{tab:S-identification-correlations}
\begin{tabular}{lr}
\toprule
Parameter pair & Correlation\\
\midrule
$\lambda_A,\sigma_a$ & -0.682058\\
$\sigma_f,\sigma_a$ & +0.505126\\
$\lambda_A,\sigma_f$ & -0.456326\\
$\sigma_a,\sigma_{rf}$ & +0.398852\\
$\sigma_f,\sigma_{rf}$ & +0.308547\\
$\lambda_A,\sigma_{rf}$ & -0.288943\\
$\lambda_A,\sigma_{ra}$ & +0.231151\\
$\sigma_a,\sigma_{ra}$ & -0.127136\\
$\sigma_f,\sigma_{ra}$ & -0.072167\\
$\sigma_{rf},\sigma_{ra}$ & +0.052354\\
\bottomrule
\end{tabular}
\begin{tablenotes}[flushleft]
\scriptsize
\item \textit{Note.} Values are from the frozen, adjudicated analysis outputs. Spans are reported on the probability scale unless explicitly identified as percentage points. M1 is the reader-only model; M2 is the shared-detectability model; M3 is the residual-dependence sensitivity.
\end{tablenotes}
\end{threeparttable}
\end{table}

The strongest tradeoffs involve the AI loading and anchor/FOV heterogeneity. Posterior concentration is therefore not interpreted as proof of unique factor decomposition.

\section{M3 residual-dependence sensitivity}

M3 applies
\[
\operatorname{logit}(p_{ic}^{\gamma})=
\operatorname{logit}(p_{ic}^{0})+\kappa_c(\gamma)+\gamma A_i,
\]
where \(\kappa_c(\gamma)\) preserves condition-specific mean future-reader positivity within each posterior draw.

Table~\ref{tab:S-residual-grid} gives representative points from the fixed residual-dependence sensitivity grid.
\begin{table}[htbp]
\centering
\setlength{\tabcolsep}{16pt}
\begin{threeparttable}
\textcolor{red}{\caption{Residual-dependence sensitivity}}
\label{tab:S-residual-grid}
\begin{tabular}{rrrr}
\toprule
$\gamma$ & $\theta_U$ & $\theta_E$ & $\Delta$\\
\midrule
-2.00 & 0.420478 & 0.409392 & -0.011086\\
-1.00 & 0.469963 & 0.463676 & -0.006287\\
-0.50 & 0.498715 & 0.495113 & -0.003602\\
-0.25 & 0.513535 & 0.511321 & -0.002213\\
0.00 & 0.528406 & 0.527602 & -0.000805\\
+0.25 & 0.543146 & 0.543762 & +0.000615\\
+0.50 & 0.557574 & 0.559612 & +0.002038\\
+1.00 & 0.584823 & 0.589677 & +0.004853\\
+2.00 & 0.629244 & 0.639348 & +0.010104\\
\bottomrule
\end{tabular}
\begin{tablenotes}[flushleft]
\scriptsize
\item \textit{Note.} Values are from the frozen, adjudicated analysis outputs. Spans are reported on the probability scale unless explicitly identified as percentage points. M1 is the reader-only model; M2 is the shared-detectability model; M3 is the residual-dependence sensitivity.
\end{tablenotes}
\end{threeparttable}
\end{table}

Across the full grid the point-estimate span is approximately 20.88 percentage points for \(\theta_U\), 23.00 percentage points for \(\theta_E\), and 2.12 percentage points for \(\Delta\). M3 is a sensitivity analysis, not a point-estimator competition and not a sharp partial-identification bound.

\section{Pattern-mixture missingness sensitivity}

A delta-adjusted pattern-mixture calculation uses the empirical reader-level fraction missing each FOV/condition as the mixture weight and shifts missing-like future-reader log odds by \(\gamma_{\rm mis}\in[-2,2]\). The observed-data posterior remains fixed.

Table~\ref{tab:S-missingness} summarizes the full and moderate pattern-mixture ranges.
\begin{table}[htbp]
\centering
\setlength{\tabcolsep}{16pt}
\begin{threeparttable}
\caption{Missingness sensitivity}
\label{tab:S-missingness}
\begin{tabular}{lrrr}
\toprule
Model & $\Delta$ point range & Point span (pp) & 95\% interval envelope\\
\midrule
M1 & [-0.003436,-0.000028] & 0.3408 & [-0.006001,0.001359]\\
M2 & [-0.002448,+0.001235] & 0.3683 & [-0.015335,0.013368]\\
\bottomrule
\end{tabular}
\begin{tablenotes}[flushleft]
\scriptsize
\item \textit{Note.} Values are from the frozen, adjudicated analysis outputs. Spans are reported on the probability scale unless explicitly identified as percentage points. M1 is the reader-only model; M2 is the shared-detectability model; M3 is the residual-dependence sensitivity.
\end{tablenotes}
\end{threeparttable}
\end{table}

At \(\gamma_{\rm mis}=0\), the calculation reproduces the original model summaries to numerical precision. The analysis does not identify the true missingness mechanism.

\section{Slide/patient hierarchy and LOSO composition sensitivity}

The slide extension adds a noncentered slide/patient random effect with \(\sigma_{\rm slide}\sim\mathrm{HalfNormal}(0.5)\).

Full-data fit:
\[
\theta_U=0.578619,\quad
\theta_E=0.578576,\quad
\Delta=-0.000043,
\]
with 95\% credible interval for \(\Delta\) [-0.013928,0.012614]. The slide-scale posterior mean was 0.638, median 0.607, equal-tail 95\% interval [0.354,1.099]. With only five slides, this is not a precise population-level variance estimate.

Table~\ref{tab:S-loso-composition} reports the five leave-one-slide-out refits.
\begin{table}[htbp]
\centering
\setlength{\tabcolsep}{5pt}
\begin{threeparttable}
\caption{Slide-composition sensitivity}
\label{tab:S-loso-composition}
\begin{tabular}{lrrrrr}
\toprule
Dropped slide & $\theta_U$ & $\theta_E$ & $\Delta$ & 95\% CrI for $\Delta$ & $P(\Delta>0)$\\
\midrule
TCGA-A1-A0SP-01Z-00-DX1 & 0.536947 & 0.541238 & +0.004291 & [-0.016292,0.025401] & 0.6614\\
TCGA-A7-A0DA-01Z-00-DX1 & 0.646438 & 0.640248 & -0.006191 & [-0.028996,0.007536] & 0.2438\\
TCGA-AR-A1AR-01Z-00-DX1 & 0.596605 & 0.594927 & -0.001679 & [-0.013332,0.008058] & 0.3925\\
TCGA-C8-A12V-01Z-00-DX1 & 0.499005 & 0.505628 & +0.006623 & [-0.019509,0.035555] & 0.6826\\
TCGA-E2-A158-01Z-00-DX1 & 0.573837 & 0.573295 & -0.000542 & [-0.015338,0.013489] & 0.4773\\
\bottomrule
\end{tabular}
\begin{tablenotes}[flushleft]
\scriptsize
\item \textit{Note.} Values are from the frozen, adjudicated analysis outputs. Spans are reported on the probability scale unless explicitly identified as percentage points. M1 is the reader-only model; M2 is the shared-detectability model; M3 is the residual-dependence sensitivity.
\end{tablenotes}
\end{threeparttable}
\end{table}

LOSO point-estimate ranges were 14.74 percentage points for \(\theta_U\), 13.46 percentage points for \(\theta_E\), and 1.28 percentage points for \(\Delta\). This is a slide-composition diagnostic, not external validation.

\begin{figure}[H]
\centering
\includegraphics[width=0.8\textwidth]{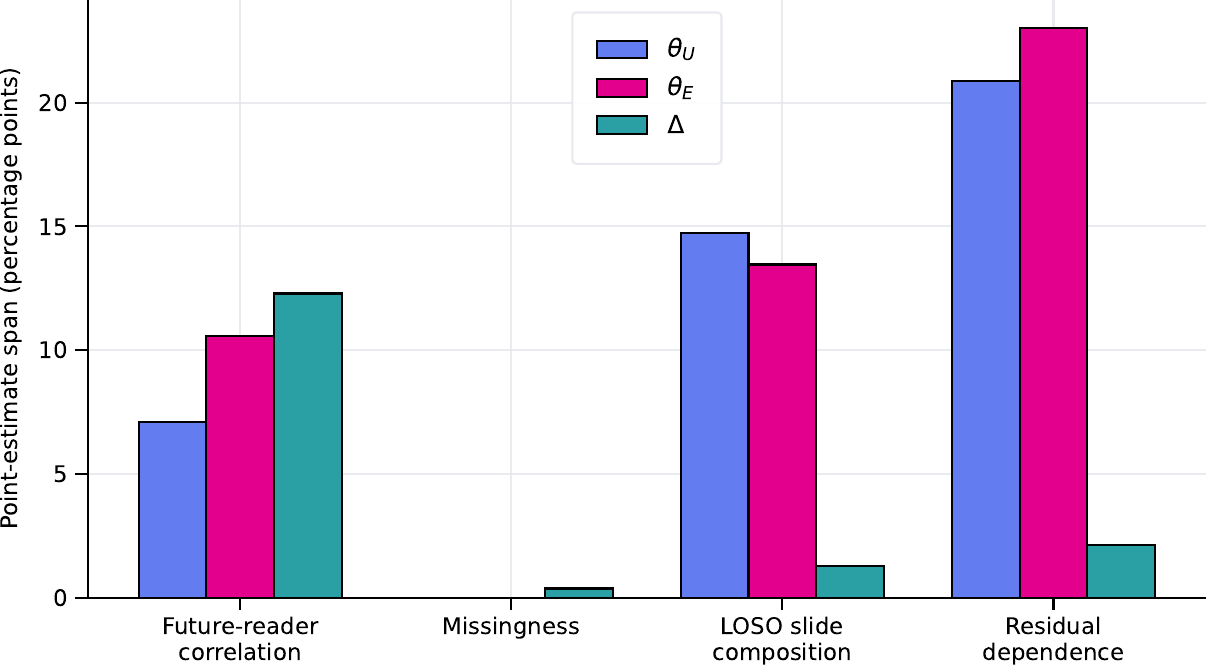}
\caption{Structural sensitivity spans}
\label{fig:S-structural-spans}
\end{figure}

Figure~\ref{fig:S-structural-spans} compares the reported point-estimate spans across the four structural sensitivity families. The missingness analysis targets the contrast only, so no absolute-target bars are shown for that family. The figure makes the central asymmetry visible: absolute agreement is most sensitive to residual dependence and slide composition, whereas the contrast is generally less variable, except under the future-reader construction range.

\section{Frozen adversarial simulation program}

\subsection{Freeze protocol and truth screen}

The frozen simulation battery consists of eight scenarios and 100 frozen replications per scenario. Fitting code cannot regenerate the data or access truth during estimation.

Table~\ref{tab:S-simulation-mechanisms} identifies the eight departures and the corresponding frozen truth-screen summaries.
\begin{table}[htbp]
\centering
\setlength{\tabcolsep}{3pt}
\begin{threeparttable}
\caption{Adversarial simulation mechanisms}
\label{tab:S-simulation-mechanisms}
\begin{tabular}{llrrrrr}
\toprule
Scenario & Departure & $n$ & AI prev. & $\theta_U$ & $\theta_E$ & $\Delta$\\
\midrule
A1 & Independent AI-specific FOV/anchor factors & 1,144 & 0.4892 & 0.4927 & 0.4930 & 0.0003\\
A2 & Correlated nonidentical AI/human factors ($\rho=0.5$) & 1,144 & 0.4902 & 0.5142 & 0.5158 & 0.0016\\
A3 & Threshold/nonlinear AI response & 1,144 & 0.5205 & 0.5523 & 0.5542 & 0.0019\\
A4 & Condition-specific AI-human dependence & 1,144 & 0.4856 & 0.5113 & 0.5388 & 0.0275\\
A5 & Omitted slide/patient effect & 1,144 & 0.4747 & 0.5393 & 0.5434 & 0.0041\\
A6 & Informative FOV missingness & 1,144 & 0.4838 & 0.5274 & 0.5304 & 0.0030\\
A7 & Selection on AI + independent human discovery read & 798 & 0.6153 & 0.6177 & 0.6166 & -0.0012\\
A8 & 90/10 mixture anchor heterogeneity & 1,144 & 0.4864 & 0.5347 & 0.5378 & 0.0031\\
\bottomrule
\end{tabular}
\begin{tablenotes}[flushleft]
\scriptsize
\item \textit{Note.} Values are from the frozen, adjudicated analysis outputs. Spans are reported on the probability scale unless explicitly identified as percentage points. M1 is the reader-only model; M2 is the shared-detectability model; M3 is the residual-dependence sensitivity.
\end{tablenotes}
\end{threeparttable}
\end{table}

\subsection{Prespecified comparative-stability rule}

\[
R=\frac{\operatorname{RMSE}(\Delta)}
{\{\operatorname{RMSE}(\theta_U)+\operatorname{RMSE}(\theta_E)\}/2}.
\]
The prespecified revision-program criterion required median \(R\le0.50\) across all 16 M1/M2 scenario cells and at least 12/16 cells with \(R<1\).

Observed:
\[
\operatorname{median}(R)=0.295447,\qquad 14/16\text{ cells have }R<1.
\]
The only \(R\ge1\) cells are M1-A4 (1.351) and M2-A4 (1.845).

\subsection{Operating characteristics and Monte Carlo uncertainty}

RMSE and bias in Table~\ref{tab:S-absolute-coverage} are percentage points. The Monte Carlo SE of the mean contrast error is derived from the reported RMSE and bias as
\[
\mathrm{MCSE}(\widehat{\mathrm{bias}})
=
\sqrt{\frac{\mathrm{RMSE}^2-\widehat{\mathrm{bias}}^2}{n-1}},
\]
and coverage MCSE is the binomial
\[
100\sqrt{\frac{\widehat p(1-\widehat p)}{n}}
\]
in percentage points.

Table~\ref{tab:S-absolute-coverage} reports empirical coverage for the two absolute agreement targets with binomial Monte Carlo standard errors.
\begin{table}[htbp]
\centering
\setlength{\tabcolsep}{16pt}
\begin{threeparttable}
\caption{Absolute-target coverage}
\label{tab:S-absolute-coverage}
\begin{tabular}{llrr}
\toprule
Scenario & Model & $\theta_U$ coverage (MCSE) & $\theta_E$ coverage (MCSE)\\
\midrule
A1 & M1 & 98.0 (1.4) & 98.0 (1.4) \\
A1 & M2 & 82.8 (3.8) & 83.8 (3.7) \\
A2 & M1 & 3.0 (1.7) & 4.0 (2.0) \\
A2 & M2 & 93.0 (2.6) & 96.0 (2.0) \\
A3 & M1 & 0.0 (0.0) & 0.0 (0.0) \\
A3 & M2 & 99.0 (1.0) & 96.0 (2.0) \\
A4 & M1 & 63.0 (4.8) & 0.0 (0.0) \\
A4 & M2 & 21.0 (4.1) & 67.0 (4.7) \\
A5 & M1 & 16.0 (3.7) & 21.0 (4.1) \\
A5 & M2 & 99.0 (1.0) & 99.0 (1.0) \\
A6 & M1 & 1.0 (1.0) & 2.0 (1.4) \\
A6 & M2 & 99.0 (1.0) & 98.0 (1.4) \\
A7 & M1 & 0.0 (0.0) & 0.0 (0.0) \\
A7 & M2 & 98.0 (1.4) & 99.0 (1.0) \\
A8 & M1 & 0.0 (0.0) & 0.0 (0.0) \\
A8 & M2 & 95.0 (2.2) & 98.0 (1.4) \\
\bottomrule
\end{tabular}
\begin{tablenotes}[flushleft]
\scriptsize
\item \textit{Note.} Values are from the frozen, adjudicated analysis outputs. Spans are reported on the probability scale unless explicitly identified as percentage points. M1 is the reader-only model; M2 is the shared-detectability model; M3 is the residual-dependence sensitivity.
\end{tablenotes}
\end{threeparttable}
\end{table}

The replication-level summaries available in the local handoff preserve absolute-target RMSE and coverage but not the corresponding absolute-target mean bias values. Therefore bias MCSEs for \(\theta_U\) and \(\theta_E\) are not reconstructed here; they should be generated directly from the canonical replication-level file during repository closeout.

A4 is the shared failure boundary: both Bayesian models have 0\% contrast coverage and approximately -2.3 to -2.5 percentage-point contrast bias. A small \(R\) elsewhere is a comparative-stability statistic, not a global model-quality score; for example, some M1 absolute-level coverages are poor even while \(\Delta\) is much more stable than the absolute levels.

\subsection{M3 adversarial envelope}

Table~\ref{tab:S-envelope-coverage} evaluates whether the M3 sensitivity envelope contains the true contrast.
\begin{table}[H]
\centering
\setlength{\tabcolsep}{8pt}
\begin{threeparttable}
\caption{M3 contrast sensitivity-envelope truth containment.}
\label{tab:S-envelope-coverage}
\begin{tabular}{lrrrr}
\toprule
Scenario & Mean span & Median span & Any-grid CI containment & Full-envelope containment\\
\midrule
A1 & 0.020620 & 0.019264 & 0.99 & 1.00\\
A2 & 0.020113 & 0.020214 & 1.00 & 1.00\\
A3 & 0.019712 & 0.018727 & 1.00 & 1.00\\
A4 & 0.033201 & 0.033097 & 0.94 & 0.94\\
A5 & 0.017753 & 0.015625 & 1.00 & 1.00\\
A6 & 0.021572 & 0.020278 & 1.00 & 1.00\\
A7 & 0.017713 & 0.016355 & 0.99 & 1.00\\
A8 & 0.018588 & 0.019027 & 1.00 & 1.00\\
\bottomrule
\end{tabular}
\begin{tablenotes}[flushleft]
\scriptsize
\item \textit{Note.} Values are from the frozen, adjudicated analysis outputs. Spans are reported on the probability scale unless explicitly identified as percentage points. M1 is the reader-only model; M2 is the shared-detectability model; M3 is the residual-dependence sensitivity.
\end{tablenotes}
\end{threeparttable}
\end{table}

The six A4 full-envelope failures had the true positive contrast above the envelope by approximately 0.10--0.68 percentage points. M3 therefore mitigates residual dependence uncertainty but does not absorb arbitrary condition-specific AI-human dependence.

\subsection{Computational adjudication}

M1 ultimately has 800/800 computationally valid fits after targeted retry of 33 divergence-only failures. M2 used 5,000 retained draws/chain, 3,000 tuning draws, four chains, and target acceptance 0.999. One M2 fit, A1 replicate 067, remained a persistent convergence failure after retry (0 divergences but R-hat up to about 1.08 and bulk ESS about 51); it is reported, not replaced or selected using truth. Final M2 validity is 799/800.

\section{Predictive validation}

\subsection{Outer-fold target and matched comparators}

The five outer folds hold out anchors while retaining all 52 FOVs and all five slides. Therefore they assess new anchors within observed contexts. For held-out AI, human observations from the same held-out anchor remain available while the AI likelihood contribution is masked; this tests shared information, not AI correctness. For held-out human labels, the target anchor is removed from the human likelihood. Target-aligned calibration evaluates \(I(A=D)\) directly.

Regularized logistic regression using reader-summary features and histogram gradient boosting were tuned within each outer training fold. Slide- and anchor-cluster bootstrap summaries were computed.

\subsection{Nested tuning}

Table~\ref{tab:S-nested-tuning} records the model and hyperparameter selected within each outer training fold.
\begin{table}[H]
\centering
\setlength{\tabcolsep}{1.0pt}
\begin{threeparttable}
\caption{Nested tuning results}
\label{tab:S-nested-tuning}
\begin{tabular}{clcl}
\toprule
Fold & Model & Inner-CV negative log-loss score & Selected parameters\\
\midrule
0 & Regularized logistic & -0.647806 & C=0.1\\
0 & Histogram gradient boosting & -0.663939 & l2=1, learning rate=0.05, max leaves=7\\
1 & Regularized logistic & -0.647467 & C=10\\
1 & Histogram gradient boosting & -0.655191 & l2=1, learning rate=0.05, max leaves=7\\
2 & Regularized logistic & -0.650586 & C=100\\
2 & Histogram gradient boosting & -0.668493 & l2=1, learning rate=0.05, max leaves=7\\
3 & Regularized logistic & -0.648545 & C=100\\
3 & Histogram gradient boosting & -0.669115 & l2=1, learning rate=0.05, max leaves=7\\
4 & Regularized logistic & -0.651845 & C=0.1\\
4 & Histogram gradient boosting & -0.666769 & l2=1, learning rate=0.05, max leaves=7\\
\bottomrule
\end{tabular}
\begin{tablenotes}[flushleft]
\scriptsize
\item \textit{Note.} Values are from the frozen, adjudicated analysis outputs. Spans are reported on the probability scale unless explicitly identified as percentage points. M1 is the reader-only model; M2 is the shared-detectability model; M3 is the residual-dependence sensitivity.
\end{tablenotes}
\end{threeparttable}
\end{table}

\subsection{Pooled predictive performance}

Table~\ref{tab:S-condition-predictive} separates pooled held-out-human and target-aligned predictive performance by annotation condition. For held-out human responses, the joint and reader-only models had similar proper scoring and discrimination in both conditions. The joint model had slightly lower Brier score and log loss in Unbiased and Evaluation, whereas differences in AUROC and AUPRC were small and did not consistently favor one model. For target-aligned agreement, both models achieved substantially higher discrimination than for direct human-response prediction, with AUROC values of approximately 0.87 in both conditions. The target-aligned Brier scores and log losses equal the corresponding held-out-human values because agreement with the fixed AI is a deterministic recoding of the binary human outcome conditional on AI status; discrimination changes because the positive target is redefined as agreement rather than human detection.

\begin{table}[H]
\centering
\setlength{\tabcolsep}{3pt}
\begin{threeparttable}
\caption{Condition-specific predictive performance}
\label{tab:S-condition-predictive}
\begin{tabular}{llrrrrrr}
\toprule
Target & Model & $N$ & Prevalence & Brier & Log loss & AUROC & AUPRC\\
\midrule
Held-out human U & Joint & 5,629 & 0.8403 & 0.1303 & 0.4253 & 0.6317 & 0.8923\\
Held-out human U & Reader-only & 5,629 & 0.8403 & 0.1309 & 0.4264 & 0.6269 & 0.8957\\
Held-out human E & Joint & 5,204 & 0.8267 & 0.1356 & 0.4353 & 0.6720 & 0.8989\\
Held-out human E & Reader-only & 5,204 & 0.8267 & 0.1365 & 0.4380 & 0.6604 & 0.8947\\
Target-aligned U & Joint & 5,629 & 0.5466 & 0.1303 & 0.4253 & 0.8733 & 0.8751\\
Target-aligned U & Reader-only & 5,629 & 0.5466 & 0.1309 & 0.4264 & 0.8746 & 0.8778\\
Target-aligned E & Joint & 5,204 & 0.5473 & 0.1356 & 0.4353 & 0.8734 & 0.8810\\
Target-aligned E & Reader-only & 5,204 & 0.5473 & 0.1365 & 0.4380 & 0.8734 & 0.8792\\
\bottomrule
\end{tabular}
\begin{tablenotes}[flushleft]
\scriptsize
\item \textit{Note.} Results are pooled over the fixed five-fold anchor-level partition. Held-out-human targets predict the observed binary reader response, whereas target-aligned targets predict whether that response agrees with the fixed AI decision. U denotes the Unbiased condition, and E denotes the Evaluation condition. Lower Brier score and log loss indicate better probabilistic prediction; higher AUROC and AUPRC indicate better discrimination. AUPRC should be interpreted relative to the prevalence of the corresponding target. These analyses retain the observed slide and FOV contexts and do not constitute new-patient or external validation.
\end{tablenotes}
\end{threeparttable}
\end{table}

\begin{figure}[htbp] \centering \includegraphics[width=0.8\textwidth]{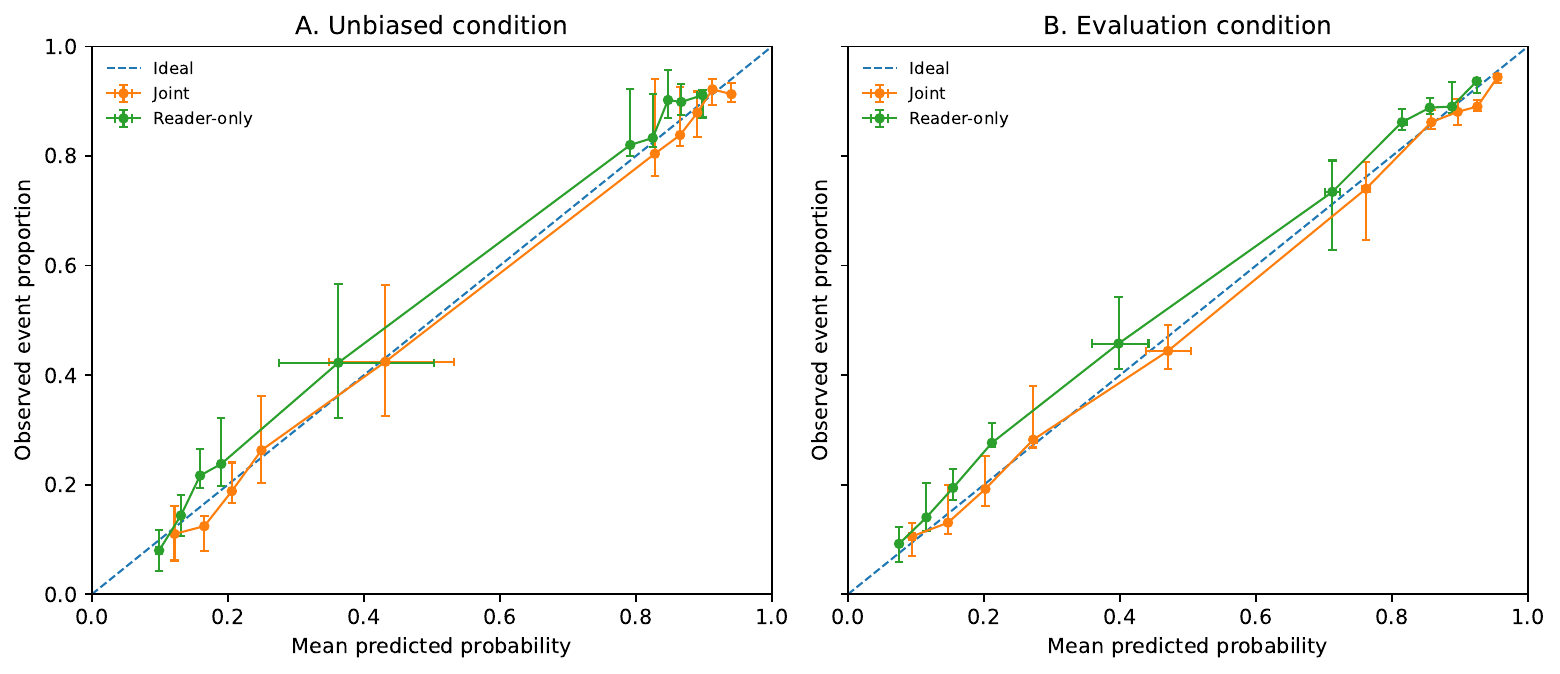} \caption{Target-aligned predictive reliability by annotation condition. Panel A shows the Unbiased condition, and Panel B shows the Evaluation condition. Points are the mean predicted probability and observed agreement proportion within quantile-based probability bins for the joint and reader-only models. Horizontal error bars are percentile 95\% slide-cluster bootstrap intervals for the mean predicted probability, and vertical error bars are percentile 95\% slide-cluster bootstrap intervals for the observed agreement proportion. The dashed diagonal denotes ideal calibration.} \label{fig:S-predictive-reliability} \end{figure}

The condition-specific results show no material predictive advantage of the joint model over the reader-only model. For held-out human responses, the joint model reduced the Brier score by 0.0006 in Unbiased and 0.0009 in Evaluation and reduced log loss by 0.0011 and 0.0027, respectively. Differences in discrimination were also small: the joint model had slightly higher AUROC in both conditions, but AUPRC was slightly lower in Unbiased and slightly higher in Evaluation. For target-aligned agreement, AUROC and AUPRC were nearly identical across models. These results indicate that incorporating the fixed AI channel did not produce a substantial or uniform improvement in out-of-fold prediction of individual human responses or AI--human agreement.

Figure~\ref{fig:S-predictive-reliability} complements the aggregate metrics by showing target-aligned reliability separately for the Unbiased and Evaluation conditions. Predicted probabilities were divided into up to 10 quantile-based bins separately for each model and condition. Each point compares the original-sample mean predicted probability with the corresponding observed agreement proportion. Horizontal and vertical error bars quantify uncertainty under resampling of the five slides.

Both models generally followed the ideal calibration line, particularly in the upper probability range where most target-aligned observations were concentrated. The reader-only model tended to lie above the diagonal in several low- and intermediate-probability bins, indicating observed agreement proportions greater than the corresponding mean predicted probabilities in those bins. The joint model was generally closer to the diagonal in that region, although the slide-cluster intervals were wide and often overlapped. At high predicted probabilities, the two models behaved similarly. Accordingly, the figure provides modest evidence of local calibration differences but does not establish uniform calibration superiority of either model. The width of several intervals also reinforces the dependence of these diagnostics on the composition of the five observed slides.

\subsection{LOSO predictive composition sensitivity}

All 15 LOSO Bayesian predictive fits were computationally clean: 0 divergences, maximum scalar R-hat 1.00, minimum bulk ESS 1,498, minimum tail ESS 1,732.

Held-out-AI metrics varied substantially by slide: Brier 0.2248--0.4271, log loss 0.6459--1.1652, AUROC 0.2988--0.6516, AUPRC 0.1798--0.8667. Held-out-slide AI prevalence ranged from 0.1437 to 0.7879.

Table~\ref{tab:S-loso-predictive} provides slide-specific target-aligned predictive diagnostics for both Bayesian models.
\begin{table}[H]
\centering
\setlength{\tabcolsep}{1pt}
\begin{threeparttable}
\scriptsize
\caption{Leave-one-slide-out predictive diagnostics}
\label{tab:S-loso-predictive}
\begin{tabular}{lrrrrrrrrrrr}
\toprule
Slide & Obs. & Joint pred. & J Brier & J logloss & J AUROC & J AUPRC & Reader pred. & R Brier & R logloss & R AUROC & R AUPRC\\
\midrule
A1 & 0.643729 & 0.674562 & 0.155247 & 0.499580 & 0.809942 & 0.855107 & 0.619892 & 0.150731 & 0.477933 & 0.816375 & 0.857682\\
A7 & 0.300803 & 0.523372 & 0.229483 & 0.650178 & 0.724080 & 0.610519 & 0.252099 & 0.148319 & 0.472314 & 0.761442 & 0.683541\\
AR & 0.363443 & 0.469901 & 0.112196 & 0.387493 & 0.896587 & 0.880126 & 0.387026 & 0.092488 & 0.331703 & 0.921853 & 0.891273\\
C8 & 0.734990 & 0.762000 & 0.131274 & 0.436629 & 0.815225 & 0.898696 & 0.694248 & 0.128635 & 0.422824 & 0.808194 & 0.898668\\
E2 & 0.644719 & 0.644425 & 0.105277 & 0.362861 & 0.884351 & 0.907696 & 0.596494 & 0.107811 & 0.374649 & 0.883480 & 0.897502\\
\bottomrule
\end{tabular}
\begin{tablenotes}[flushleft]
\scriptsize
\item \textit{Note.} Values are from the frozen, adjudicated analysis outputs. Spans are reported on the probability scale unless explicitly identified as percentage points. M1 is the reader-only model; M2 is the shared-detectability model; M3 is the residual-dependence sensitivity.
\end{tablenotes}
\end{threeparttable}
\end{table}

These are target-specific slide-composition diagnostics. They do not establish zero-information new-slide or new-patient validation.

\begin{figure}[htbp]
\centering
\includegraphics[width=0.8\textwidth]{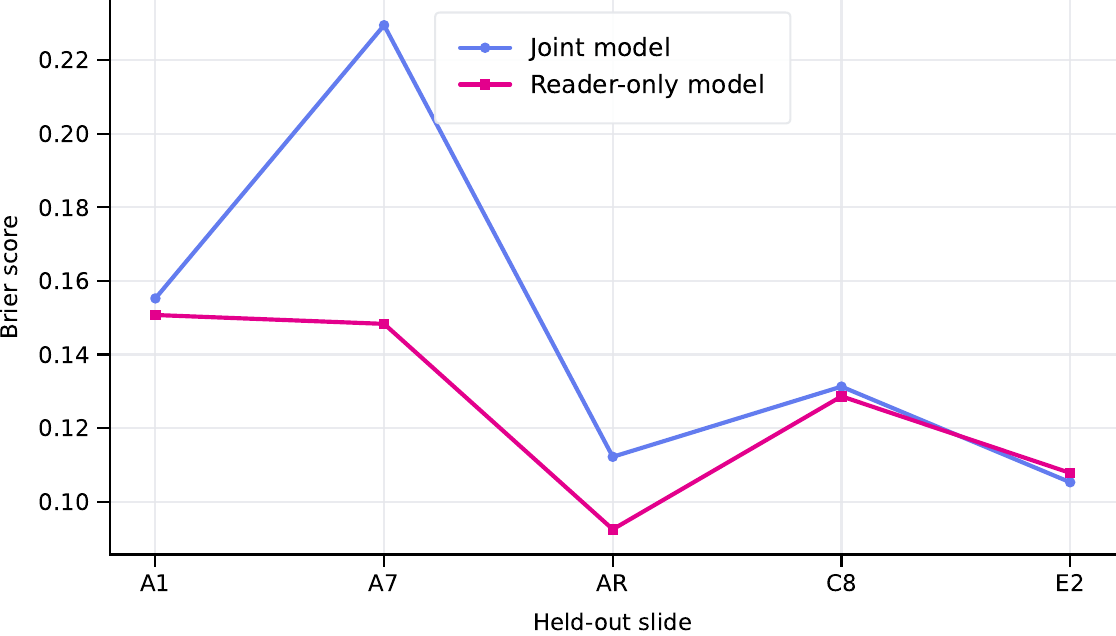}
\caption{Leave-one-slide-out Brier scores}
\label{fig:S-loso-brier}
\end{figure}

Figure~\ref{fig:S-loso-brier} displays the slide-specific Brier scores from Table~\ref{tab:S-loso-predictive}. The joint and reader-only models are similar on four slides, while the A7 holdout shows a substantially larger joint-model Brier score. This pattern reinforces that pooled predictive summaries do not remove sensitivity to slide composition.

\section{Software environment and reproducibility}
The definitive Python environment used Python 3.11.12 with PyMC 5.28.5, PyTensor 2.38.3, ArviZ 0.23.4, NumPy 2.4.6, pandas 3.0.5, SciPy 1.17.1, scikit-learn 1.9.1, and xarray 2026.7.0. The repository contains analysis scripts, environment specifications, frozen simulation materials, and derived outputs. Computational-platform, scheduler, job, and storage-path details are intentionally omitted because they are not part of the scientific method.

\section{Cross-validation assignment}
The held-out-AI assessment used a fixed five-fold partition at the anchor level. The 1,144 selected anchors appeared in exactly one outer test fold, and all observations associated with a test anchor inherited that anchor's fold assignment. Model fitting, preprocessing, and hyperparameter selection were restricted to the corresponding outer training data. For the held-out-AI target, the AI likelihood contribution for the test anchor was masked while same-anchor human responses remained available; the estimand therefore quantifies prediction of the fixed AI decision from human-response information within the observed slide and FOV contexts. For the held-out-human target, the target response was omitted from the human likelihood. The separate LOSO analysis held out each slide intact and therefore assesses composition sensitivity rather than zero-information transport to a new patient population. Table~\ref{tab:S-nested-tuning} documents the inner-loop choices, while Table~\ref{tab:S-loso-predictive} and Figure~\ref{fig:S-loso-brier} document the intact-slide diagnostic.